\documentclass[amsmath,amssymb,nofootinbib,reprint,superscriptaddress,pra,aps]{revtex4-1}

\usepackage{graphicx}
\usepackage{graphics}
\usepackage{dcolumn}
\usepackage{bm}
\usepackage{color}
\usepackage{verbatim}
\usepackage{soul,xcolor}
\usepackage{rotating}
\usepackage{setspace}
\usepackage[mathlines]{lineno}
\usepackage{mathcomp}
\usepackage{verbatim}
\DeclareMathAlphabet{\mathpzc}{OT1}{pzc}{m}{it}
\usepackage{physics}
\usepackage{amsmath}
\usepackage[10pt]{moresize}
\usepackage[caption=false]{subfig}
\usepackage{microtype}

\def\expv#1{\left\langle #1 \right\rangle}

\def\bok#1#2#3{\left\langle#1\left|#2\right|#3 \right\rangle}

\def\ket#1{|\,#1\, \rangle}
\def\braket#1{\langle\,#1\,\rangle}
\def\TB{{\textnormal{\tiny PB}}}
\def\BB{{\textnormal{\tiny BB}}}

\newcommand{\GFTyMA}{
Grupo de F{\'i}sica Te{\'o}rica y Matem{\'a}tica Aplicada, Instituto de F{\'i}sica, 
Facultad de Ciencias Exactas y Naturales, 
Universidad de Antioquia; Calle 70 No. 52-21, Medell\'in, Colombia.}
\newcommand{\GFAM}{Grupo de F\'i­sica At\'omica y Molecular, Instituto de F\'isica, 
Facultad de Ciencias Exactas y Naturales, Universidad de Antioqu­ia UdeA; 
Calle 70 No. 52-21, Medell\'i­n, Colombia.}
\newcommand{\FICOMACO}{Grupo de F{\'i}sica Computacional en Materia Condensada, Escuela de F{\'i­}sica, Facultad de Ciencias, Universidad Industrial de Santander; Cra 27 Calle 9 Ciudad Universitaria, Bucaramanga, Colombia.}

\usepackage[hyperindex=true,pdftitle={Nonadiabatic Sunlight-harvesting},colorlinks=true,pagebackref=false,citecolor=blue,plainpages=false,pdfpagelabels,linkcolor=blue,urlcolor=blue]{hyperref}

\begin{document}

\title{Nonadiabatic Sunlight-harvesting}

\affiliation{\GFTyMA}
\author{Leonardo F. Calder\'on}
\affiliation{\FICOMACO}
\affiliation{\GFTyMA}

\author{Leonardo A. Pach\'on}
\email{leonardo.pachon@udea.edu.co}
\affiliation{\GFTyMA}
\affiliation{\GFAM}

\begin{abstract} 
Experimental and theoretical evidence point out to the crucial role of specific resonant 
intramolecular vibrational modes in the interpretation of long-lived coherences observed 
in two-dimensional spectra of some natural and synthetic light harvesting complexes.
For the natural situation of illumination by incoherent (sun)light, the relevance of 
these vibrations is analyzed here for light-harvesting vibronic prototype dimers.
The detailed analysis of the density matrix dynamics reveals that  
the inclusion of the intramolecular vibrational modes reinforce up to one order of
magnitude the exciton coherence
and may increase the populations of lowest energy single exciton states, 
as well as populations and coherences in the site basis.
In sharp contrast to the case of initial-state preparation by coherent 
(laser)light-sources, the initial thermal state of the local vibrational modes, as well 
as that of the anti-correlated mode, evolves devoid of non-classical correlations 
as confirmed by the absence of negative values of its phase-space quasi--probability 
distribution at all times.
Therefore, not only the long-lived coherences observed in two-dimensional spectra are 
induced by the coherent character of pulsed laser sources, but it is unambiguously shown 
here that the non-classical character generally assigned to the anti-correlated 
vibrational mode also comes as the result of the preparation of the initial state  by
coherent pulsed laser sources.
\end{abstract}

\date{\today}
\maketitle

In the last three decades, new experimental, theoretical and computational 
techniques have been developed to resolve the interplay between the multiple electronic
and vibrational degrees of freedom, together with the variety of energy scales involved in 
molecular-aggregates energy-transfer-processes.
\cite{PachonBrumer2012PCCP,ChenuScholes2015ARPC,BrunkRothlisberger2015CR,CurutchetMennucci2016CR,deVegaAlonso2017RMP,JangMenucci2018RMP,Brumer2018JPCL}. 
In doing so, two-dimensional-electronic-spectroscopy experiments have revealed long-lived 
oscillations in two-dimensional spectra of several photosynthetic light-harvesting complexes \cite{EngelCalhounReadEtAl2007N,ColliniWongWilkEtAl2010N,PanitchayangkoonHayesFranstedEtAl2010PNAS} 
that pointed out to the potential existence of non-trivial quantum effects related to the 
interplay of the electronic and vibrational degrees of freedom \cite{IshizakiCalhounSchlau-CohenEtAl2010PCCP,PachonBrumer2011JPCL,PachonBrumer2012PCCP,HuelgaPlenio2013CP,ChenuScholes2015ARPC}.  
In particular, the coupling between the electronic degrees of freedom and intramolecular
vibrations in quasi-resonance to excitonic transitions (vibronic coupling) has been proposed 
as a consistent physical design principle that could explain the origin of long-lived oscillations 
observed in two-dimensional spectra, and possibly related to the high efficiency 
of the energy transfer process 
\cite{ChristenssonKauffmannPulleritsEtAl2012JPCB,KolliOReillyScholesEtAl2012JCP,TiwariPetersJonas2013PNASU,ChinPriorRosenbachEtAl2013NP,ChenuChristenssonKauffmannEtAl2013SR,NovelliNazirRichardsEtAl2015JPCL,MalySomsenNovoderezhkinEtAl2016CPC,DeanMirkovicToaEtAl2016C,YehHoehnAllodiEtAl2018PNAS}.
Unexpectedly, during the course of potentially being supporting long-lived oscillations, the state of 
the intramolecular vibrations evolve from a thermal state with non-quantum correlations 
into a state provided with genuinely quantum correlations even at room temperature \cite{OReillyOlaya-Castro2014NC}.

Two-dimensional-electronic-spectroscopy is a laser-pulsed non-linear technique 
\cite{Mukamel:1995}
and therefore, the extend to which their results are representative for natural conditions with continuous 
incoherent light sources has been intensively addressed in the literature \cite{MancalValkunas2010NJP,BrumerShapiro2012PNASU,TscherbulBrumer2014PRL,SadeqBrumer2014JCP,GrinevBrumer2015JCP,DodinTscherbulBrumer2016JCP,PachonBoteroBrumer2017JPBAMOP,Brumer2018JPCL,ChenuChristenssonKauffmannEtAl2013SR,ChenuMalyMancal2014CP,ChenuBranczykScholesEtAl2015PRL,ChenuBrumer2016JCP}.
It is by now clear that the dynamics induced by suddenly-turned-on incoherent sunlight 
are qualitatively different from coherent laser sources and also different from a bare
white-noise-source provided that incoherent light has a super-Ohmic character and 
does not induce pure dephasing dynamics 
\cite{PachonBrumer2013PRA,PachonBoteroBrumer2017JPBAMOP}.
The relevance of intramolecular vibrational modes, their impact on the energy transfer, 
as well as their non-trivial quantum character are explored here for the natural scenario
of illumination by sunlight.

To account for that natural scenario in which chromophores harvest incoherent 
sunlight in the presence of localized vibrational modes, consider vibronic dimers formed
by chromophores treated within the two-level approximation and a quantized 
intramolecular vibrational mode in interaction with each monomer. 
The vibronic dimers are initially set up in their electronic ground and thermal 
vibrational state.
The excitation ignited by sunlight is then dissipated by a thermal phonon bath that 
accounts for the effects of the protein and solvent environments. 
The subsequent dynamics of the vibronic dimer are compared, in the site and exciton bases, with 
the dynamics of its corresponding electronic dimer with no specific intramolecular 
vibrational modes.  
As concrete examples, the two phycoerythrobilin chromophores from 
the protein-antenna phycoerythrin 545 and the two dihydrobiliverdin chromophores from the
protein-antena phycocyanin 645 of marine cryptophyte algae are considered below.

Under sunlight illumination, it is found that the introduction of intramolecular vibrational 
modes may increase the population amplitudes in the vibronic dimers compared to the 
electronic ones.
The amplitude of the single exciton coherence increases up to one order of magnitude 
with the inclusion of the intramolecular vibrational modes, but coherence between site 
states is of the same order for both electronic and vibronic dimers.
To account for the assistance of intramolecular vibrational modes to long-decoherence
times,
the decoherence rate of vibronic single exciton state superpositions is systematically 
analyzed in a broad regime of the dimer parameter space.
It is shown that the decoherence rate displays a non-trivial behavior and that neither 
the nonadiabatic regime nor the values of electronic and vibronic couplings of the 
dynamics analyzed lead to the lowest value in the decoherence rate.

The fact that the dynamics induced by the \textit{ suddenly-turning-on} of incoherent radiation are 
effectively coherent, in the vibronic single exciton basis, may equivocally leads to the
conclusion that
the temporal coherence of the light source plays a minor role on the energy-transfer
process.
By focusing on the non-trivial quantum characteristics of the states of intramolecular vibrational 
degrees of freedom, it is ambiguously shown that under sunlight illumination conditions, the
state of the two vibrational modes and that of the anticorrelated vibrational mode evolve devoid 
of non-classical correlations.
This is in sharp contrast to the case of illumination by coherent light sources \cite{OReillyOlaya-Castro2014NC}
and can be considered as a genuine and experimentally verifiable difference between 
natural and in-lab conditions, \textit{independent of the suddenly-turning-on condition}.

\section{Light-harvesting antenna systems under sunlight 
\label{sec:theory1}}
Consider a molecular aggregate immersed within a protein that is excited by incoherent 
sunlight radiation. 
The global system-bath Hamiltonian reads $\hat{H} = \hat{H}_{\mathrm{S}} + \hat{H}_{\mathrm{SB}} 
+ \hat{H}_{\mathrm{B}}$.
The light-harvesting system corresponds to a set of $N$ chromophores (sites) whose
electronic degrees of freedom are coupled to quantized intramolecular vibrational modes
and described by the Hamiltonian 
\begin{equation}
\label{equ:system_hamiltonian}
\begin{split}
\hat{H}_{\mathrm{S}} = &
\sum_{i=1}^{N}\left(E_{\textnormal{g}_i}\hat{\textnormal{1}}_i + \epsilon_{i}\hat{\sigma}_{i}^{+}
\hat{\sigma}_{i}^{-}\right)
+ \sum_{i \ne j}^{N}V_{ij}\hat{\sigma}_{i}^{+}\hat{\sigma}_{j}^{-} %
\\&
+ \sum_{i=1}^{N}\hbar\mathpzc{g}_{i}\hat{\sigma}_{i}^{+}\hat{\sigma}_{i}^{-}\left(\hat{\mathpzc{b}}_{i}^{\dag}+\hat{\mathpzc{b}}_{i}\right) 
+ \sum_{i=1}^{N}\hbar\varpi_{i}\hat{\mathpzc{b}}_{i}^{\dag}\hat{\mathpzc{b}}_{i},
\end{split}
\end{equation}
being $E_{\textnormal{g}_i}$  the ground state energy, $\epsilon_{i}$ the 
electronic energy of the $i^\mathrm{th}$ site, $\hat{\sigma}_{i}^{+}$ ($\hat{\sigma}_{i}^{-}$) creates 
(annihilates) an electronic excitation in the $i^\mathrm{th}$ site, $V_{ij}$ is the
electronic coupling between the $i^\mathrm{th}$
and the $j^\mathrm{th}$ site. 
Here, $\hat{\mathpzc{b}}_{i}^{\dag}$ ($\hat{\mathpzc{b}}_{i}$) is the creation (annihilation) 
operator of the $i^\mathrm{th}$ intramolecular vibrational mode of frequency $\varpi_{i}$,
and
$\mathpzc{g}_i=\sqrt{S_i}\varpi_i$ represents the coupling between the $i^\mathrm{th}$
excited eletronic state and the $i^\mathrm{th}$ intramolecular vibrational mode, and 
$S_i$ is the Huang-Rhys factor.

The action of the operators $\hat{\sigma}_{i}^{+}$ and $\hat{\mathpzc{b}}_{i}^{\dag}$ 
allows for defining, in an adiabatic basis, the vibronic states (i.e.,
electronic-vibrational states) in the electronic ground, singly excited, and doubly
excited states through the relations
$
\ket{g_i,\nu_i} = \left(\hat{\mathpzc{b}}_{i}^{\dag}\right)^{\nu_i}/\sqrt{\nu_i!}\ket{g_i,0}, 
$
$
\ket{\epsilon_i,\nu_i} = \hat{\sigma}_{i}^{+} \left(\hat{\mathpzc{b}}_{i}^{\dag}\right)^{\nu_i}/\sqrt{\nu_i!}\ket{g_i,0}, 
$
$
\ket{f_{ij},\nu_i} = \hat{\sigma}_{i}^{+}\hat{\sigma}_{j}^{+}\left(\hat{\mathpzc{b}}_{i}^
{\dag}\right)^{\nu_i}/\sqrt{\nu_i!}\ket{g_i,0} ,
$
respectively. $\nu_i$ stands for the vibrational quantum number of the $i^\mathrm{th}$ site and
$\ket{g_i,0}$ for the electronic-vibrational ground state. 
The eigenstates of the Hamiltonian $\hat{H}_{\mathrm{S}}$ correspond to the vibronic
exciton states $\{\ket{\psi_n}\}$ defined by $\hat{H}_{\mathrm{S}}\ket{\psi_n}=\xi_n
\ket{\psi_n}$. 
The vibronic single exciton states are defined by
$
\ket{\psi_n} = \sum_{i=1}^{N}\sum_{\nu_i} C_{i,\nu_i}^{n} \ket{\epsilon_i,\nu_i}.
$
The localization of the $n^\mathrm{th}$ vibronic single exciton state on the 
$k^\mathrm{th}$ site is given by 
$l_{\epsilon_k}(\psi_n) = \braket{\psi^k_n | \psi_n} =  \sum_{\nu_k}\left|C_{k,\nu_k}^{n}\right|^2$, 
being
$\ket{\psi^k_n} = \sum_{\nu_k} C_{k,\nu_k}^{n} \ket{\epsilon_k,\nu_k}$
a vibronic single exciton state completely localized on the $k^\mathrm{th}$ site. 
The intersite mixing ratio for a superposition between the $n^\mathrm{th}$ and the 
$m^\mathrm{th}$ vibronic single exciton states of a vibronic dimer is defined by  
\cite{MalySomsenNovoderezhkinEtAl2016CPC} 
$
\zeta_{nm} = 
l_{\epsilon_1}(\psi_n)l_{\epsilon_2}(\psi_m) +
l_{\epsilon_1}(\psi_m)l_{\epsilon_2}(\psi_n).
$
This ratio characterizes the type of coherence  in the  vibronic single exciton basis:
$\zeta_{nm}=1$ for a pure electronic coherence (superposition between vibronic single
exciton states localized each one on different sites), and $\zeta_{nm}=0$ for a
pure vibrational coherence (superposition between vibronic single exciton states
localized on the same site).

\begin{figure*}
\includegraphics[scale=1.0]{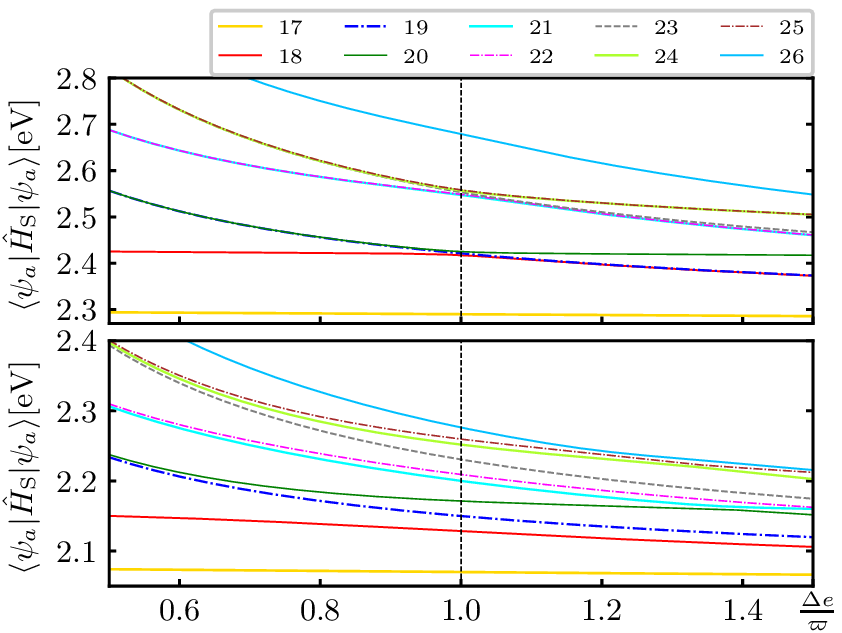}
\includegraphics[scale=1.0]{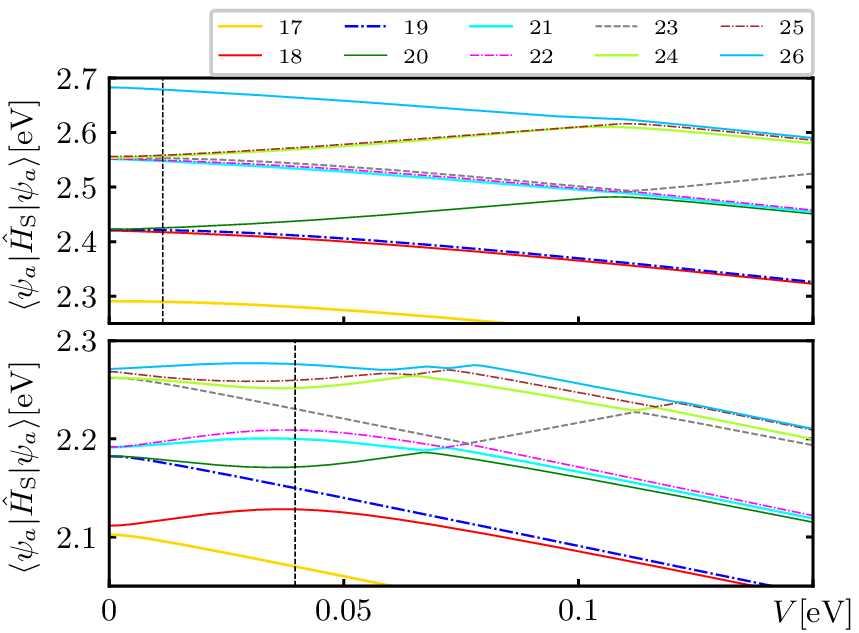}
\caption{\label{fig:PEB-DBV_EnergyLandscape} Left panels: $\langle\psi_a|\hat{H}_{\mathrm{S}}|\psi_a\rangle$ as a function of the ratio between the exciton energy splitting $\Delta e$ and the intramolecular vibrational frequency $\varpi$. Right panels: $\langle\psi_a|\hat{H}_{\mathrm{S}}|\psi_a\rangle$ as a function of the electronic coupling $V$. For both panels, upper figures: PEB dimer, and bottom figures: DBV dimer. The vertical black dashed lines indicate the conditions considered in the simulations.}
\end{figure*}

Due to the exponential scaling of the dimension of the full Hilbert space, for all simulations 
below, only the first four states (ground state and three excited levels)  of 
each intramolecular vibrational mode are considered. 
Convergence of the density matrix time evolution was verified with the case of four 
excited levels.
For the vibronic dimers considered here (two monomers and two intramolecular
vibrations), the vibronic exciton manifold has then dimension 64: 16 
vibronic ground exciton states $\{\ket{\psi_1},\dots,\ket{\psi_{16}}\}$, 32 
vibronic single exciton states $\{\ket{\psi_{17}},\dots,\ket{\psi_{48}}\}$ and 16 
vibronic double exciton states $\{\ket{\psi_{49}},\dots,\ket{\psi_{64}}\}$. 
The comparison with the case of an electronic dimer with no specific intramolecular 
vibrational modes, in the site $\{\ket{\epsilon_1},\ket{\epsilon_2}\}$ and in the exciton 
basis $\{\ket{e},\ket{e^{\prime}}\}$, follows after tracing over the intramolecular 
vibrational degrees of freedom in the density matrix of the vibronic dimer.
For electronic dimers, the Frenkel Hamiltonian corresponds to the first two terms of the 
Hamiltonian described in Eq.~(\ref{equ:system_hamiltonian}). 
Thus, the two monomers have a site representation described by the states 
$\{\ket{\tilde{\epsilon}_1},\ket{\tilde{\epsilon}_2}\}$ and by two single exciton states 
$\{\ket{\tilde{e}},\ket{\tilde{e}^{\prime}}\}$. 

Specifically, the two phycoerythrobilin ($\mathrm{PEB}_{50/61\;\mathrm{C}}$ and $
\mathrm{PEB}_{50/61\;\mathrm{D}}$) chromophores from the protein-antenna
phycoerythrin 545 (PE545), and the two dihydrobiliverdin ($\mathrm{DBV}_{50/61\;\mathrm{C}}$
and $\mathrm{DBV}_{50/61\;\mathrm{D}}$) chromophores from the
protein-antena phycocyanin 645 (PC645) of marine cryptophyte algae are considered below.
The PEB dimer has a large energy gap between excited electronic states 
$\Delta\epsilon=1042\;\mathrm{cm}^{-1}$, and due to the 
large spatial separation between chromophores, the electronic coupling is small 
($V=92\;\mathrm{cm}^{-1}$); in consequence, each excitonic state is highly localized 
over a specific chromophore. 
The exciton energy splitting is $\Delta e=1058\;\mathrm{cm}^{-1}$.
The DBV dimer has a small energy gap between excited electronic states
$\Delta\epsilon=73\;\mathrm{cm}^{-1}$, and moderate
electronic coupling ($V=319.4\;\mathrm{cm}^{-1}$) that results in the formation of
delocalized exciton states with
an exciton energy splitting $\Delta e=643\;\mathrm{cm}^{-1}$.
Figure~\ref{fig:PEB-DBV_EnergyLandscape} depicts the functional dependence of 
$\langle\psi_a|\hat{H}_{\mathrm{S}}|\psi_a\rangle$ on the ratio between the exciton 
energy splitting and the intramolecular vibrational frequency $\Delta e/\varpi$, 
and on the electronic coupling $V$ for the first ten vibronic single exciton states 
($\ket{\psi_{17}},\dots,\ket{\psi_{26}}$) for 
the PEB (top panels) and DBV (bottom panels) dimers. 
The vertical dashed lines indicate the conditions considered in the simulations below, 
and corresponds to a nonadiabatic framework 
\cite{TiwariPetersJonas2013PNASU,YehHoehnAllodiEtAl2018PNAS}
that has been related to an enhancement of energy transfer process and the
appearance of non-trivial quantum correlations driven by strong vibronic interactions 
\cite{OReillyOlaya-Castro2014NC,NovelliNazirRichardsEtAl2015JPCL,ScholesFlemingChen2017N}.

The environment of the protein complex in Eq.~(\ref{equ:system_hamiltonian})
can be treated as a local phonon bath, whereas sunlight is formally described 
as blackbody radiation at $5600~K$.
Due to the highly mixed character of electronic-vibrational coherences, it is then 
necessary also to consider the coupling of the vibrational modes to the environment 
of the protein complex.
The system-bath Hamiltonian is given by 
\begin{equation}
\begin{split}
\hat{H}_{\mathrm{SB}} = & 
\sum_{i,l}^{N,\infty} \hbar g_{il}^{(\mathrm{e})} \hat{\sigma}_{i}^{+}\hat{\sigma}_{i}^{-}
\left( \hat{b}^{(i)}_{l} + {\hat{b}_{l}}^{(i)\dag}  \right) 
- \sum_{j}^{N} \hat{\boldsymbol{\mu}}_j\cdot \hat{\mathbf{E}}(t)
\\ &
+ \sum_{i,m}^{N,\infty} \hbar g_{im}^{(\mathrm{v})} (\hat{\mathpzc{b}}_{i}^{\dag} + \hat{
\mathpzc{b}}_{i})
\left( \hat{b}^{(i)}_{m} + {\hat{b}_{m}}^{(i)\dag}  \right) ,
\label{eq:system_baths_hamiltonian} 
\end{split}
\end{equation}
\begin{equation}
\begin{split}
\hat{H}_{\mathrm{B}}  = &
\sum_{i,l}^{N,\infty} \hbar \omega_l^{(i)} \hat{b}_{l}^{(i)\dag} \hat{b}_{l}^{(i)} 
+ \sum_{\mathbf{k},s} \hbar c k \hat{a}_{\mathbf{k},s}^{\dag} \hat{a}_{
\mathbf{k},s} \\&
+ \sum_{i,m}^{N,\infty} \hbar \omega_m^{(i)} \hat{b}_{m}^{(i)\dag} \hat{b}_{m}^{(i)} . 
\label{eq:baths_hamiltonian}
\end{split}
\end{equation}
Here, $g_{il}^{(\mathrm{e})}$ ($g_{im}^{(\mathrm{v})}$) represent the coupling between
the $i^\mathrm{th}$ site ($i^\mathrm{th}$ intramolecular vibrational mode) and 
the $l^\mathrm{th}$ ($m^\mathrm{th}$) phonon mode. 
$\hat{b}_l^{(i)\dag}$ $\big(\hat{b}_l^{(i)}\big)$ is the
creation (annihilation) operator of a $l^\mathrm{th}$ phonon mode of frequency 
$\omega_l^{(i)}$ which interacts with the $i^\mathrm{th}$ site.
$\hat{b}_m^{(i)\dag}$ $\big(\hat{b}_m^{(i)}\big)$ is the
creation (annihilation) operator of a $m^\mathrm{th}$ phonon mode of frequency 
$\omega_m^{(i)}$ which interacts with the $i^\mathrm{th}$ intramolecular vibrational mode. 
$\hat{\boldsymbol{\mu}}_j$ is the dipole operator for the $i^\mathrm{th}$ site and 
the electric field of the radiation \cite{Mandel-Wolf:1995} is given by 
$\hat{\mathbf{E}}(t)= \hat{\mathbf{E}}^{(+)}(t)+ \hat{\mathbf{E}}^{(-)}(t)$, with 
$ \hat{\mathbf{E}}^{(+)}(t)= \mathrm{i} \sum_{\mathbf{k},s} 
\left(\frac{\hbar \omega}{2 \epsilon_0 V}\right)^{1/2} \hat{a}_{\mathbf{k},s} (\varepsilon_{\mathbf{k},s})
\mathrm{e} ^{-\mathrm{i} \omega t}$ and $ \hat{\mathbf{E}}^{(-)}(t)=\left[\hat{\mathbf{E}}^{(+)}(t)\right]^\dagger$, 
and $\hat{a}_{\mathbf{k},s}^{\dag}$ ($\hat{a}_{\mathbf{k},s} $) being the creation 
(annihilation) operator for the $\mathbf{k}^\mathrm{th}$ radiation field mode in the
$s^\mathrm{th}$ polarization state.

\subsection{System dynamics\label{sec:theory2}}

\begin{figure*}
\includegraphics[scale=1.0]{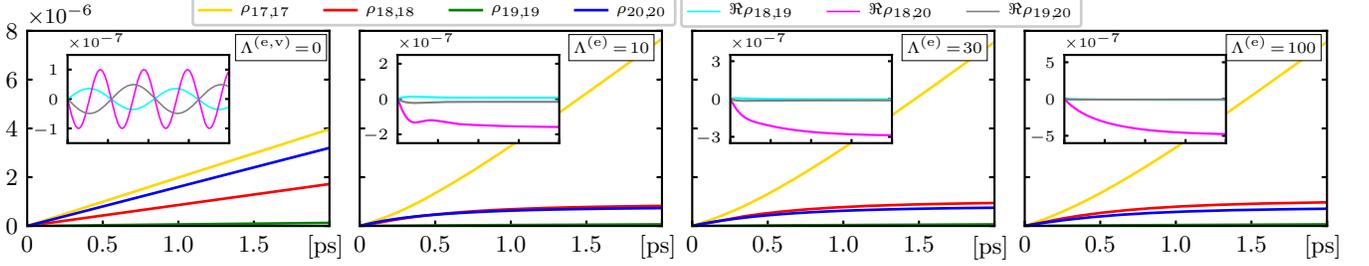}
\caption{\label{fig:PEB_rhovibronic_popcoh_TBBB} 
Vibronic single exciton states populations $\rho_{aa}=\bok{\psi_a}{\hat{\rho}}{\psi_a}$ 
and coherences (inset figures)
$\mathfrak{R}\rho_{ab}=\mathfrak{R}\bok{\psi_a}{\hat{\rho}}{\psi_b}$ (color coding is
shown on top) in the PEB dimer for 
the reorganization energies
$\Lambda^{(\mathrm{e})}=0,10,30,100\;\mathrm{cm}^{-1}$ and $\Lambda^{(\mathrm{v})}=10 \;\mathrm{cm}^{-1}$.
Baths parameters are $T_
{\TB}^{(\mathrm{e},\mathrm{v})}=300\,\mathrm{K}$,
$T_{\BB}=5600\,\mathrm{K}$.}
\end{figure*}

The effect of the incoherent radiation 
environment (blackbody bath) and the vibrational environment (phonon bath) is encoded
in the spectral densities $J^{\BB}(\omega) $ and $J_{j}^{\TB}(\omega)$, respectively.
The blackbody radiation bath is characterized by the super-Ohmic spectral density
\cite{PachonBoteroBrumer2017JPBAMOP} 
\begin{equation}
\label{eq:spectral_density_blackbody_bath}
\omega^{2}J_{j}^{\BB}(\omega) = \frac{2\hbar\omega^3}{3(4\epsilon_0\pi^2c^3)}.
\end{equation}
This spectral density generates long-lasting coherent dynamics (see Fig.~\ref{fig:PEB_rhovibronic_popcoh_TBBB}, 
$\Lambda^{(\mathrm{e},\mathrm{v})}=0$ case) provided by the lack of pure dephasing dynamics and the strong 
dependence of the decoherence rate on the system level spacing \cite{PachonBoteroBrumer2017JPBAMOP} 
(see Fig.~\ref{fig:PEB_gamma1820}~A and B). 
The spectral density of the phonon-baths reads
\begin{equation}
\label{eq:spectral_density_thermal_bath}
\omega^{2}J_{j}^{\TB}(\omega) = \frac{2\Omega_{j}^{(\mathrm{e},\mathrm{v})}\Lambda_{j}^{(\mathrm{e},
\mathrm{v})}\omega}{\hbar(\omega^{2}+\Omega_{j}^{(\mathrm{e},\mathrm{v})2})},
\end{equation}
where $\Omega^{(\mathrm{e},\mathrm{v})}$ represents the cutoff frequency and 
$\Lambda^{(\mathrm{e},\mathrm{v})}$ the reorganization energy of the phonon baths
coupled to the electronic (e) and intramolecular vibrational (v) degrees of freedom.
The dynamics of light-harvesting systems with spectral densities of the form 
(\ref{eq:spectral_density_thermal_bath}) are commonly solve with the hierarchical 
equations of motion (HEOM) method 
\cite{TanimuraKubo1989JPSJ,IshizakiTanimura2005JPSJ,IshizakiFleming2009JCP}.
However, this method cannot be applied to super-ohmic spectral densities.
To circumvent this, the HEOM method has been used to treat the
phonon bath whereas the non-unitary effect of the incoherent light has been accounted 
for by a Lindblad dissipator \cite{FassioliOlaya-CastroScholes2012JPCL,ChanGamelFlemingEtAl2018JPB}. 
However, this hybrid approach does not properly get into account the influence of the
super-Ohmic spectral density in the density matrix dynamics  since it does not
consider the dependence of the decoherence rates on the system level spacing.
To adequately describe the correlations induced by the super-Ohmic spectral 
density of the blackbody radiation together with the phonon bath effects, 
the dynamics are solved in the vibronic exciton basis $\{|\psi_n \rangle\}$,  
i.e., in the eigenstates of the Hamiltonian in Eq.~(\ref{equ:system_hamiltonian}),
by using the standard Redfield master equation (second-order and non-secular) 
\cite{May-Kuhn:2011,PachonBoteroBrumer2017JPBAMOP}.
Recent works on energy transfer dynamics of vibronic dimers excited with coherent light 
have considered the Redfield approach and have shown similar results to the HEOM method 
under parameters used in experimental conditions
\cite{RomeroAugulisNovoderezhkinEtAl2014NP,NovoderezhkinGrondelle2017JPB,BennettMalyKreisbeckEtAl2018JPCL}.
Further details about both methods and comparison between them are
presented in {{\it SI}, Sec. 1 and 2}.

\section{Dynamics in the presence of blackbody radiation and phonon baths 
\label{sec:dynamics_onlyBB}}

Energy transfer starts with the rapid incoherent excitation of the electronic 
sites in their electronic ground states with transition dipole moments of $11.87\;
\mathrm{D}$ ($\mathrm{PEB}_{50/61\;\mathrm{D}}$), $12.17\;\mathrm{D}$ 
($\mathrm{PEB}_{50/61\;\mathrm{C}}$), 
$13.1\;\mathrm{D}$ ($\mathrm{DBV}_{50/61\;\mathrm{D}}$) 
and $13.2\;\mathrm{D}$ ($\mathrm{DBV}_{50/61\;\mathrm{C}}$).
Initially, each chromophore is in its ground electronic state and the intramolecular
vibrational modes are in thermal equilibrium at $T=300\,\mathrm{K}$. After the
dynamics begin, the vibronic dimer remains coupled to the incident blackbody
radiation \cite{PachonBrumer2013PRA,PachonBoteroBrumer2017JPBAMOP}; this is in sharp
contrast to the pulsed laser excitation conditions 
\cite{Brumer2018JPCL}.
The frequency of the two intramolecular vibrational modes is in full resonance with 
the exciton splitting, i.e., $\varpi_{1}=\varpi_{2}=\Delta e$. The vibronic coupling
strength to each monomer is the same, i.e., $\mathpzc{g}_1=\mathpzc{g}_2=\mathpzc{g}$,
specifically, $\mathpzc{g}=267.1\;\mathrm{cm}^{-1}$ for the PEB dimer and 
$\mathpzc{g}=250\;\mathrm{cm}^{-1}$ for the DBV dimer.
For the simulations below, same spectral densities on each monomer are taken, 
with $\Omega_j^{(\mathrm{e})}=100\;\textnormal{cm}^{-1}$ and for various values of the
reorganization energy 
$\Lambda_j^{(\mathrm{e})}$. 
Besides, same spectral densities on each intramolecular vibrational mode are taken, with 
$\Omega_j^{(\mathrm{v})}=50\;\textnormal{cm}^{-1}$ and reorganization energy 
$\Lambda_j^{(\mathrm{v})}=10 \;\textnormal{cm}^{-1}$ provided that
$\Lambda_j^{(\mathrm{e})}\neq0$.

\subsection{Vibronic single exciton basis}
Consider first the dynamics of the vibronic dimer in interaction with blackbody radiation 
only, i.e., set to zero the non-unitary effects related to the phonon baths and assume 
that the system and the blackbody radiation are initially decoupled, i.e., 
$\hat{\rho}(t_0)=\hat{\rho}_{\mathrm{S}}(t_0)\otimes \hat{\rho}_{\mathrm{BB}}(t_0)$. 
The transition dipole moments are considered parallel to the incident electric field and 
constant in time so that the effect of different orientations of the transition 
dipole moments is neglected.

\begin{figure*}
\includegraphics[scale=1.0]{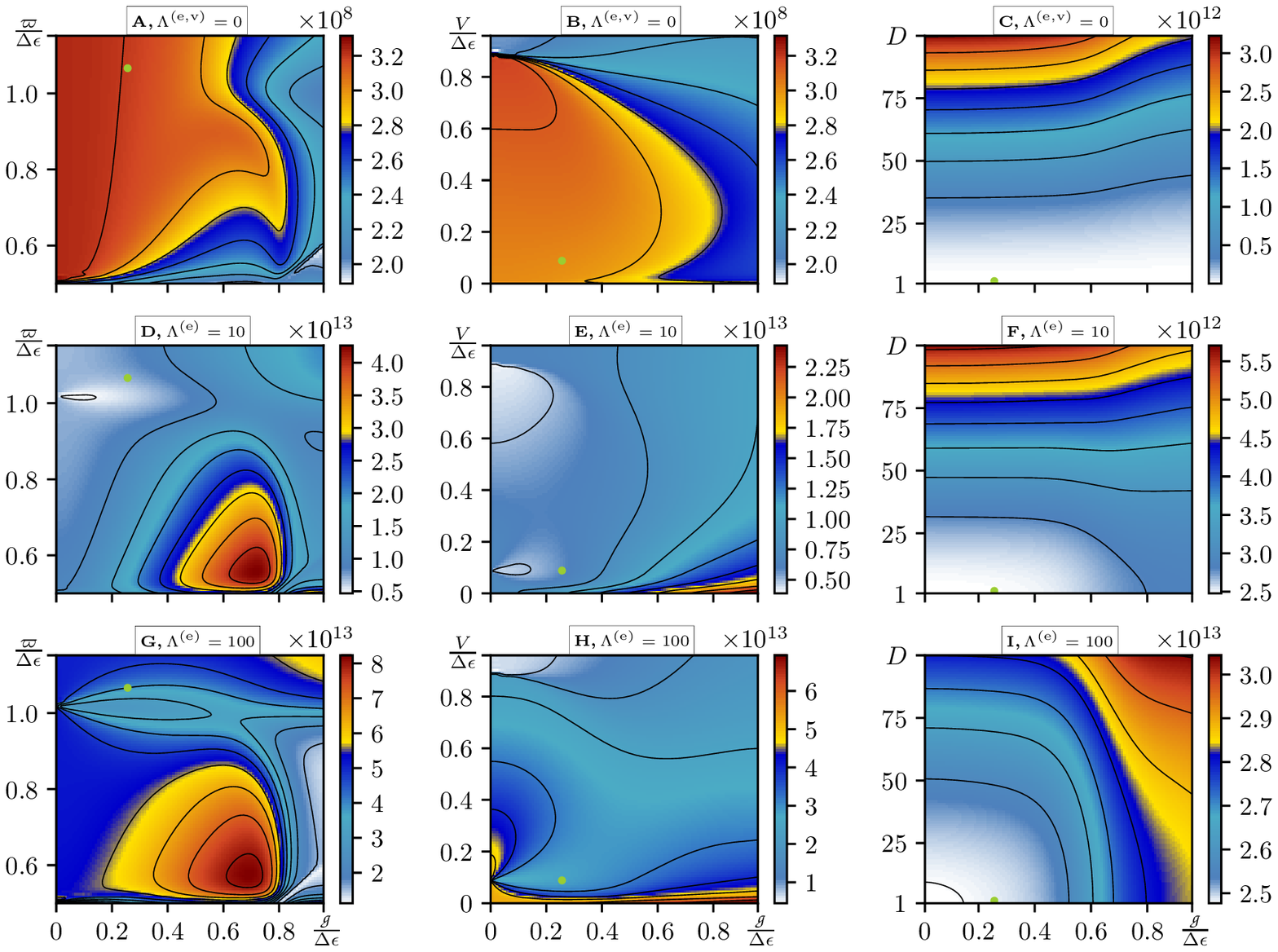}
\caption{\label{fig:PEB_gamma1820} 
Decoherence rate $\gamma_{18,20}\,[\mathrm{s}^{-1}]$ (color map) for the PEB dimer with
the reorganization energies $\Lambda^{(\mathrm{e},\mathrm{v})}=0\,\mathrm{cm}^{-1}$ (top
panels), $\Lambda^{(\mathrm{e})}=10\,
\mathrm{cm}^{-1}$, $\Lambda^{(\mathrm{v})}=10 \;\mathrm{cm}^{-1}$ (middle panels) and
$\Lambda^{(\mathrm{e})}=100\,\mathrm{cm}^{-1}$, $\Lambda^{(\mathrm{v})}=10 \;
\mathrm{cm}^{-1}$ (bottom panels), as
a function of the ratios $\mathpzc{g}/\Delta\epsilon$,  $\varpi/\Delta\epsilon$,
$V/\Delta\epsilon$ and the transition dipole moment amplitude $D$, where $\Delta\epsilon$
represents the site energy difference. 
Green points represent the values adopted for the simulations of the density matrix
dynamics in the vibronic single exciton, exciton and sites bases, discussed throughout the
paper. 
Baths parameters are $T_{\TB}^{(\mathrm{e},\mathrm{v})}=300\;\mathrm{K}$,\ $T_{\BB}=5600\;
\mathrm{K}$.}
\end{figure*}

Figure~\ref{fig:PEB_rhovibronic_popcoh_TBBB} ($\Lambda^{(\mathrm{e},\mathrm{v})}=0$ case)
shows the populations
($> 10^{-6}$)
of the vibronic single exciton states and the coherent superpositions arising between 
them in the PEB dimer after suddenly-turning-on the incoherent radiation.
The linear increase of the populations is expected in low-intensity incoherent radiation 
\cite{PachonBoteroBrumer2017JPBAMOP}. 
In chromophores isolated from the vibrational phonon environment, suddenly turned-on 
incoherent-light-induced-dynamics are effectively coherent and last for hundreds of 
picoseconds \cite{PachonBoteroBrumer2017JPBAMOP}. 
Nevertheless, the amplitude of the vibronic coherences ($\sim10^{-8}$) is approximately 
two orders of magnitude smaller than the populations; hence, they turn out to quickly 
become irrelevant for the dynamics of populations \cite{SadeqBrumer2014JCP}. 
Most of these vibronic coherences display a highly mixed electronic-vibrational 
character, quantified through the intersite mixing ratio. 
Specifically, for the coherences depicted in Fig.~\ref{fig:PEB_rhovibronic_popcoh_TBBB}, 
$\zeta_{18,19}=0.52$, $\zeta_{18,20}=0.50$ and $\zeta_{19,20}=0.48$. 
Therefore, vibronic coherence dynamics are influenced by the decoherence and dissipation 
of the electronic as well as intramolecular degrees of freedom.
This is the reason for introducing a thermal bath of each intramolecular vibrational mode in 
Eq.~(\ref{eq:baths_hamiltonian}).

To incorporate the effect of the phonon bath ($\Lambda^{(\mathrm{e},\mathrm{v})}\neq0$),
assume that it is initially decoupled from the dimers 
$\hat{\rho}(t_0) = \hat{\rho}_{\mathrm{S}}(t_0)\otimes \hat{\rho}_{\mathrm{BB}}(t_0)\otimes
\hat{\rho}_{\mathrm{PB}}(t_0)$.
Figure~\ref{fig:PEB_rhovibronic_popcoh_TBBB} depicts the populations of vibronic single 
exciton states for different values of the reorganization energy 
$\Lambda^{(\mathrm{e})}=10,30,100\,\mathrm{cm}^{-1}$. 
Specifically, the population of the lowest energy vibronic single exciton state $\ket{\psi_{17}}$
of the PEB dimer increases for increasing values of the reorganization energy.
In the case of the DBV dimer, there is an increase of two orders of magnitude in the
population of the state $
\ket{\psi_{17}}$, see
{{\it SI}, Fig.~S2}.
This is a consequence of the intricate interplay between bath-enhanced rates and
non-adiabatic dynamics of vibronic single exciton states with small energy gaps.
Figure~\ref{fig:PEB_rhovibronic_popcoh_TBBB} also depicts the dynamics of superpositions
between vibronic single exciton states (vibronic coherences). 
The vibronic coherences, originated by the turning-on of the incoherent radiation, decay 
due to the interaction with the phonon bath.
Their influence on the population of the vibronic single exciton states is negligible, 
owing to the amplitude of the vibronic coherences is approximately one (PEB dimer) and 
two (DBV dimer) orders of magnitude smaller than the populations of vibronic single
exciton states.

To explore the role of the vibronic coupling and the incoherent light excitation process 
on the lifetime of quantum superpositions between vibronic single exciton states, 
the decoherence rate of the highest amplitude vibronic coherence $\mathfrak{R}\rho_{18,20}(t)$ 
in the PEB dimer is analyzed (see Fig.~\ref{fig:PEB_rhovibronic_popcoh_TBBB}). 
Figure~\ref{fig:PEB_gamma1820} depicts the functional dependence of the decoherence rate 
$\gamma_{18,20}$ on the ratio between the vibronic coupling and the site energy difference 
$\mathpzc{g}/\Delta\epsilon$, the ratio between the intramolecular vibrational 
frequency and the site energy difference $\varpi/\Delta\epsilon$, the ratio between the
electronic coupling and the site energy difference $V/\Delta\epsilon$, and the 
dipole moment amplitude $D$.
The green points in Fig.~\ref{fig:PEB_gamma1820} depict the specific values for the PEB dimer,
$\mathpzc{g}/\Delta\epsilon=0.26$, $\varpi/\Delta\epsilon= 1.02$, $V_{12}/\Delta\epsilon=0.09$ and $D=1$ 
(For the DBV dimer case, see {{\it SI}, Fig.~S3}).

In absence of thermal baths for the intra-molecular vibrational modes (not shown), the 
decoherence rate $\gamma_{18,20}$ decreases for increasing values of the vibronic coupling 
$\mathpzc{g}$.
However, in the more realistic scenario depicted in Fig.~\ref{fig:PEB_gamma1820}, increasing
the vibronic coupling may lead to regions of parameter space with higher decoherence rates.
Thus, the decoherence rate  $\gamma_{18,20}$ displays a non-trivial behavior under the 
variation of the physical quantities defined above and neither the vibronic resonance condition 
$\varpi_{1}=\varpi_{2}=\Delta e$ (i.e., nonadiabatic regime)  nor the values of electronic and 
vibronic couplings of the dynamics discussed (see green points in Fig.~\ref{fig:PEB_gamma1820}) 
lead to the lowest value in the decoherence rate.
Thus, the longest decoherence time for vibronic single exciton state superpositions is 
not reached under the physical conditions considered usually in
two-dimensional-electronic-spectroscopy studies 
\cite{YehHoehnAllodiEtAl2018PNAS,DuanThorwartMiller2019arXiv}.

For natural light-matter coupling strengths (Figs.~\ref{fig:PEB_gamma1820} A, B, D, E, G and H):
(i) increasing the reorganization energy increases the decoherence rate $\gamma_{18,20}$ for
 $\Lambda^{(\mathrm{e})}=0,10,100\,[\mathrm{cm}^{-1}]$ as $\gamma_{18,20}\sim10^{8},10^{13},10^{13}\,[
 \mathrm{s}^{-1}]$,
respectively.
(ii) For $\Lambda^{(\mathrm{e})}=0$, the decoherence rate $\gamma_{18,20}$ is at least five orders of 
magnitude smaller than for cases with $\Lambda^{(\mathrm{e})}\neq0$.
This follows from the low intensity of sunlight and the energy-gap dependence of blackbody 
radiation decoherence rates 
$
\gamma_{e,e'}^{\BB} \sim \left(\mu_{e,e'}^2 \omega_{e,e'}^3 / 3\hbar\pi\epsilon_0 c^3 \right)
\coth(\hbar\omega_{e,e'}/2k_{\mathrm{B}}T^{\BB})
$.
Thus, for small energy gaps $\hbar\omega_{e,e'}$, the decoherence rate $\gamma_{e,e'}^{\BB} $ 
may be considerably smaller than the case of an Ohmic thermal phonon bath 
$ \gamma_{e,e'}^{\BB} \sim 4 k_\mathrm{B} T^\mathrm{PB} \Lambda / \hbar^2 \Omega$,
which is energy-gap independent \cite{PachonBoteroBrumer2017JPBAMOP}.
For the values of the reorganization energy considered in Fig.~\ref{fig:PEB_gamma1820}, 
the increase of the transition dipole moment amplitude  leads to higher values in the 
decoherence rate $\gamma_{18,20}$ (see Fig.~\ref{fig:PEB_gamma1820} C, F and I).
Increasing of the vibronic coupling $\mathpzc{g}$ in Fig.~\ref{fig:PEB_gamma1820} for the values of
the reorganization energy considered does not leads to decrease of the decoherence rate.

\subsection{Reduced exciton and site bases}
\begin{figure*}
\includegraphics[scale=1.0]
{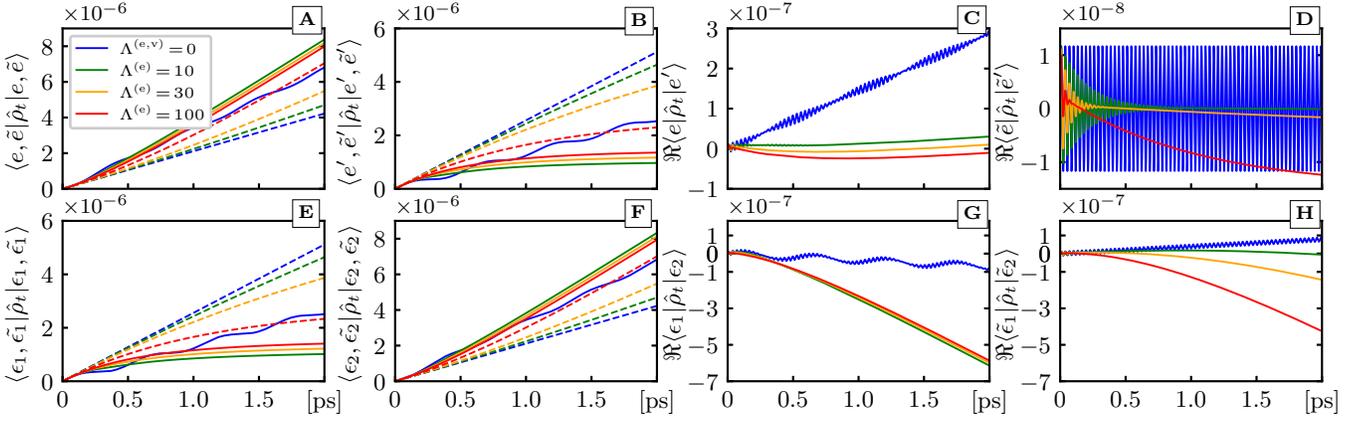}
\caption{\label{fig:PEB_rhoexcsite_popcoh_TBBB}
Top panels\textemdash Dynamics in the single exciton basis for the vibronic dimer
case $\{\ket{e},\,\ket{e'}\}$, and the electronic dimer case $\{\ket{\tilde{e}},\,\ket{
\tilde{e}'}\}$ varying the reorganization energy $\Lambda^{(\mathrm{e})}\,[\mathrm{cm}^{-1}]$ 
($\Lambda^{(\mathrm{v})}=10 \;\mathrm{cm}^{-1}$) in the PEB
dimer (color
coding is shown on the top left box): A,B) Populations of the lowest and highest energy
single exciton states (solid and dashed lines represent the vibronic and electronic dimer cases, respectively). C) Single exciton coherence in the vibronic dimer model. D) Single exciton coherence in the electronic dimer model.
Bottom panels\textemdash Dynamics in the site basis 
$\{\mathrm{PEB}_{50/61\;\mathrm{D}},\mathrm{PEB}_{50/61\;\mathrm{C}}\}$
(vibronic dimer case $\{
\ket{\epsilon_{1}},\,\ket{\epsilon_{2}}\}$, and electronic dimer $\ket{\tilde{\epsilon}_
{1}},\,\ket{\tilde{\epsilon}_{2}}$) varying the reorganization energy $\Lambda^{(\mathrm{e})}\,[
\mathrm{cm}^{-1}]$ ($\Lambda^{(\mathrm{v})}=10 \;\mathrm{cm}^{-1}$) in the PEB
dimer (same color coding as the top panels): E,F) Populations of the two 
site states in the vibronic dimer (solid lines) and the electronic dimer 
(dashed lines). G) Coherence between site states in the vibronic dimer model. 
H) Coherence between site states in the electronic dimer model.
Baths parameters are $T_{\TB}^{(\mathrm{e},\mathrm{v})}=300\,\mathrm{K}$, $T_{\BB}=5600\,\mathrm{K}$.
}
\end{figure*}
\begin{figure*}
\includegraphics[scale=1.0]
{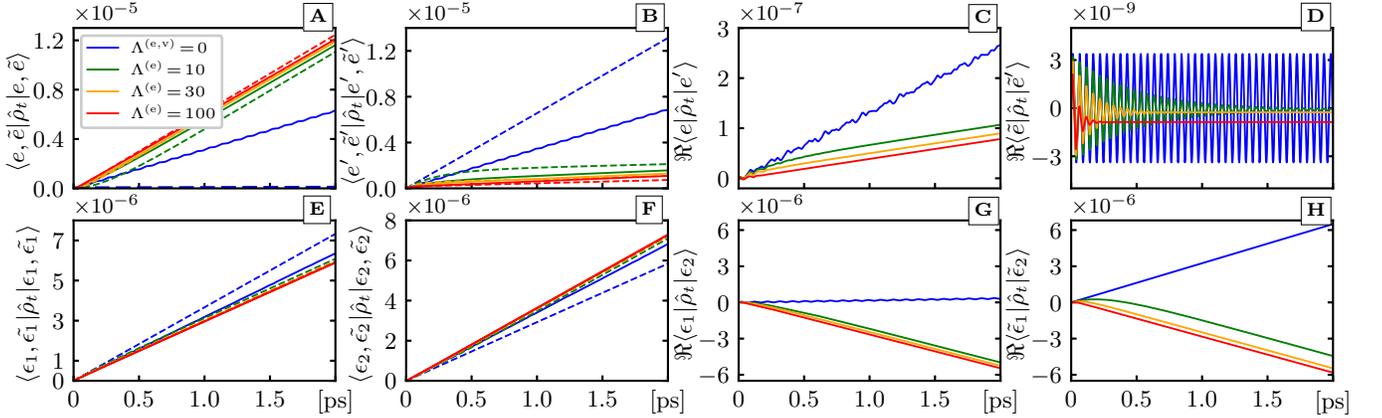}
\caption{
\label{fig:DBV_rhoexcsite_popcoh_TBBB} 
Top panels\textemdash Dynamics in the single exciton basis for the vibronic dimer
case $\{\ket{e},\,\ket{e'}\}$, and the electronic dimer case $\{\ket{\tilde{e}},\,\ket{
\tilde{e}'}\}$ varying the reorganization energy $\Lambda^{(\mathrm{e})}\,[\mathrm{cm}^{-1}]$ 
($\Lambda^{(\mathrm{v})}=10 \;\mathrm{cm}^{-1}$) in the DBV
dimer (color
coding is shown on the top left box): A,B) 
Populations of the lowest and highest energy single exciton states (solid and dashed 
lines represent the vibronic and electronic dimer cases), respectively. C) Single exciton 
coherence in the vibronic dimer model. D) Single exciton coherence in the electronic 
dimer model.
Bottom panels\textemdash Dynamics in the site basis 
$\{\mathrm{DBV}_{50/61\;\mathrm{D}},\mathrm{DBV}_{50/61\;\mathrm{C}}\}$
(vibronic dimer case $\{\ket{
\epsilon_{1}},\,\ket{\epsilon_{2}}\}$, and electronic dimer $\{\ket{\tilde{\epsilon}_
{1}},
\,\ket{\tilde{\epsilon}_{2}}\}$) varying the reorganization energy $\Lambda^{(\mathrm{e})}\,[\mathrm{cm}^
{-1}]$ ($\Lambda^{(\mathrm{v})}=10 \;\mathrm{cm}^{-1}$) in the DBV
dimer (same color coding as the top panels): E,F) Populations of the two site states in the
 vibronic dimer (solid lines) and 
the electronic dimer (dashed lines).  G) Coherence between site states in the vibronic 
dimer model. H) Coherence between site states in the electronic dimer model.
Baths parameters are $T_{\TB}^{(\mathrm{e},\mathrm{v})}=300\,\mathrm{K}$, $T_{\BB}=5600\,\mathrm{K}$.
}
\end{figure*}

The role of the high frequency intramolecular vibrational modes is explored by comparing 
the reduced electronic dynamics (tracing over the intramolecular vibrations) of the vibronic 
dimers considered above with their corresponding electronic dimer dynamics (no specific 
intramolecular vibrational modes).
Figures~\ref{fig:PEB_rhoexcsite_popcoh_TBBB} and \ref{fig:DBV_rhoexcsite_popcoh_TBBB}
show the populations and coherences in
the exciton (top panels) and site bases (bottom panels), with (vibronic dimer) 
and without (electronic dimer) intramolecular vibrational modes for the PEB and DBV
dimers, respectively.

In the exciton basis, and in the absence of the phonon baths  
($\Lambda^{(\mathrm{e},\mathrm{v})}=0\ \mathrm{cm}^{-1}$), the population of the lowest
energy exciton state
of the vibronic dimer is higher than that of its corresponding electronic dimer case: 
one and half times for the PEB dimer (see Fig.~\ref{fig:PEB_rhoexcsite_popcoh_TBBB}~A), 
and two orders of magnitude for the DBV dimer (see Fig.~\ref{fig:DBV_rhoexcsite_popcoh_TBBB}~A). 
The amplitude of the coherence between single exciton states in the vibronic dimer model 
(see Fig.~\ref{fig:PEB_rhoexcsite_popcoh_TBBB}~C) is one order of magnitude higher than 
in the electronic dimer case (see Fig.~\ref{fig:PEB_rhoexcsite_popcoh_TBBB}~D) for the PEB
dimer, and two orders of magnitude in the case of DBV dimer (see 
Fig.~\ref{fig:DBV_rhoexcsite_popcoh_TBBB}~C and D).
The increase in population and coherence can be understood as the result of the smaller energy 
gaps \cite{PachonBrumer2011JPCL,PachonBrumer2012PCCP} induced by intramolecular vibrations, 
i.e., a consequence of the non-adiabatic character of the dynamics (see Fig.~\ref{fig:PEB-DBV_EnergyLandscape}). 
Hence, for $\Lambda^{(\mathrm{e},\mathrm{v})}=0$, the population of the lowest energy exciton state and the
coherence between single exciton states increase with the inclusion of intramolecular
vibrational modes.

\begin{figure*}[!]
\includegraphics[scale=1.0]{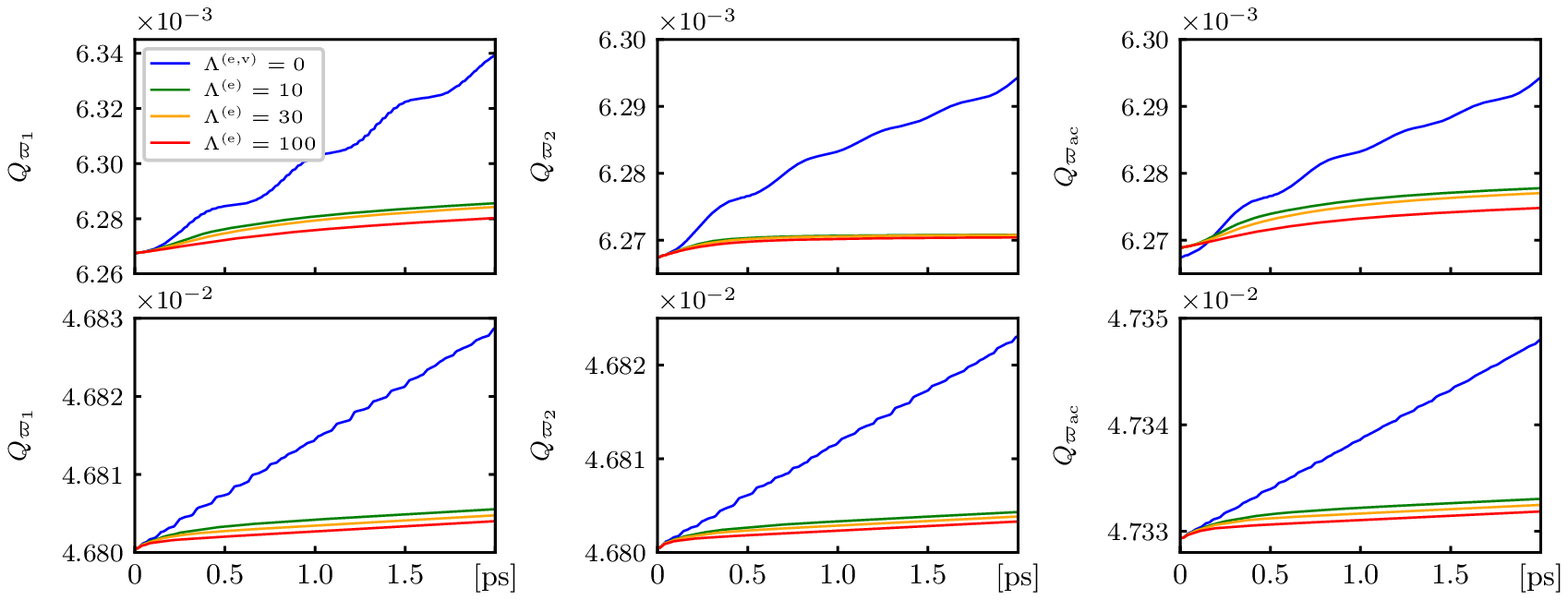}
\caption{\label{fig:peb_Mandel_parameters}Values of the Mandel parameter $Q$ for the 
vibrational modes of frequencies $\varpi_1$, $\varpi_2$, and $\varpi_{\mathrm{ac}}$ (top
pannels: PEB dimer, bottom panels: DBV dimer), for
different values of the
reorganization energies $\Lambda^{(\mathrm{e})}\,[\mathrm{cm}^{-1}]$ and $\Lambda^{(
\mathrm{v})}=10 \;\mathrm{cm}^{-1}$ (color coding
is shown on the top left).
Baths parameters are $T_{\TB}^{(\mathrm{e},\mathrm{v})}=300\,\mathrm{K}$, $T_{\BB}=5600\,\mathrm{K}$.
}
\end{figure*}
In the presence of the phonon bath ($\Lambda^{(\mathrm{e},\mathrm{v})}\neq0$), the behavior of populations are
similar to $\Lambda^{(\mathrm{e},\mathrm{v})}=0$.
The amplitude of the coherences increases slightly in the PEB dimer 
(see Fig.~\ref{fig:PEB_rhoexcsite_popcoh_TBBB}~C and D) and by up to one order 
of magnitude in the DBV dimer (see Fig.~\ref{fig:DBV_rhoexcsite_popcoh_TBBB}~C and D). 
The population of the lowest energy exciton state of the PEB dimer have higher 
amplitudes in the vibronic dimer than in the electronic dimer; however, as the value of the
reorganization increases, this population difference decreases.
Moreover, the population rate of the reduced vibronic dimer barely changes by increasing
the reorganization energy, thus seeing to be robust against the fluctuations of the
phonon environment
(see Figs.~\ref{fig:PEB_rhoexcsite_popcoh_TBBB}~A and \ref{fig:DBV_rhoexcsite_popcoh_TBBB}~A). 
For small values of the reorganization energy ($\Lambda^{(\mathrm{e})}\sim10\mathrm{cm}^{-1}$), a
similar
population 
trend is found for the DBV dimer; however, for moderate ($\Lambda^{(\mathrm{e})}\sim30\mathrm{cm}^{-1}$) and large 
values ($\Lambda^{(\mathrm{e})}\sim100\mathrm{cm}^{-1}$) of the reorganization energy, the population of the
lowest energy 
exciton state has slightly higher amplitudes in the case of the electronic dimer than the vibronic 
dimer (see Fig.~\ref{fig:DBV_rhoexcsite_popcoh_TBBB}~A).
This clearly shows the highly non-trivial interplay between bath-enhanced population rates and
non-adiabatic dynamics for highly localized (PEB) and highly delocalized (DBV) excitons.

In the site basis, and for all values of the reorganization energy considered, 
the population of the chromophores 
$\mathrm{PEB}_{50/61\;\mathrm{C}}$ and $\mathrm{DBV}_{50/61\;
\mathrm{C}}$ is higher than that of
 $\mathrm{PEB}_{50/61\;\mathrm{D}}$ and $\mathrm{DBV}_{50/61\;\mathrm{D}}$, 
respectively, in the case of the vibronic dimer than in the electronic dimer case
(see Fig.~\ref{fig:PEB_rhoexcsite_popcoh_TBBB}~E and F, and 
Fig.~\ref{fig:DBV_rhoexcsite_popcoh_TBBB}~E and F).
Since population is pumped from the ground state and the transition dipole moments
favor excitation to singly excited exciton states with low energy, the population 
difference occurs due the localization $l_{\epsilon_i}$ of the vibronic single exciton states
on those chromophores; specifically, 
$l_{\epsilon_1}(\psi_{17})=0.007,l_{\epsilon_2}(\psi_{17})=0.993$,  
$l_{\epsilon_1}(\psi_{18})=0.524,l_{\epsilon_2}(\psi_{18})=0.476$,  
$l_{\epsilon_1}(\psi_{19})=0.007,l_{\epsilon_2}(\psi_{19})=0.993$, and  
$l_{\epsilon_1}(\psi_{20})=0.475,l_{\epsilon_2}(\psi_{20})=0.525$. 
Here, $\mathrm{PEB}_{50/61\;\mathrm{D}}$ and $\mathrm{DBV}_{50/61\;\mathrm{D}}$ are
sites 1 whereas $\mathrm{PEB}_{50/61\;\mathrm{C}}$ and $\mathrm{DBV}_{50/61\;\mathrm{C}}$
are sites 2.

For the PEB dimer, the amplitude of the coherence between site states of the vibronic 
dimer is higher than that of its corresponding electronic dimer case. Also, the coherence
between site states decreases with the coupling to the phonon bath and shows to be robust
against the fluctuations of the phonon environment.
For the DBV dimer, the coherence between site states increases with the coupling to the
phonon bath and remains of the same order for electronic and vibronic dimers.
Therefore, for both dimers considered here, and for the highest values of the
reorganization energy, quantized vibrational modes barely enhance populations or
coherences in the site basis.

\section{On the quantum character of the intramolecular quantized vibrations}

It is well established that intramolecular quantized vibrational modes
initially in a thermal state could develop a genuinely non-classical character due to coherent 
exciton-vibration interactions \cite{OReillyOlaya-Castro2014NC}. 
The detailed analysis of the potential generation of non-classicality, in the context of
incoherent light excitation, allows for concluding that the quantized vibrational modes
do not display non-classical correlations quantified by the Mandel parameter \cite{Mandel-Wolf:1995}.

The Mandel parameter identifies the non-classical character of bosonic states 
through the comparison of occupation number distribution for a given bosonic state with
the occupation number distribution of a coherent state \cite{Mandel-Wolf:1995}. 
The Mandel parameter is given by 
$
Q = \left(\expv{\hat{n}^2} - \expv{\hat{n}}^2 \right)/\expv{\hat{n}} - 1
$
, where $\hat{n}$ is the occupation number operator of the bosonic state. 
For the case of a coherent state the occupation number distribution corresponds to a Poisson 
distribution ($Q=0$). 
Thus, for any occupation number distribution narrower than a Poisson distribution, i.e., with 
$Q<0$, the associated bosonic state has a quantum character with no classical analog.

For different values of the reorganization energy and under sunlight illumination conditions, 
Fig.~\ref{fig:peb_Mandel_parameters} depicts the Mandel parameter for the two vibrational 
modes of frequencies $\varpi_1$ and $\varpi_2$ considered in the vibronic model of the 
PEB and DBV dimers. 
The reduced dynamics of the anticorrelated vibrational mode of frequency 
$\varpi_{\mathrm{ac}}$, previously analyzed in the seminal work by Jonas \textit{et al.} 
\cite{TiwariPetersJonas2013PNASU} and responsible for the non-adiabatic character
of the dynamics (see Fig.~\ref{fig:PEB-DBV_EnergyLandscape}), is also considered.
For every case considered, the Mandel parameter adopted positive values, which indicates 
that during the dynamics, the state of the intramolecular vibrations modes has a classical
character.
Even, in the case of absence of phonon bath ($\Lambda^{(\mathrm{e},\mathrm{v})}=0$), the
value of the Mandel parameter remains positive. 

Under sunlight illumination conditions, dimers are initially in their the electronic ground state 
while intramolecular vibrations, that are decoupled from the ground state, are initially
at thermal equilibrium;
therefore, the initial condition is devoid of quantum superpositions.
Under pulsed-laser-excitation conditions, vibrations are also assumed to be at thermal equilibrium;
however, in sharp contrast to natural conditions, the dimer is assumed to have been prepared in,
e.g., an exciton state.
Therefore, under pulsed-laser-excitation conditions, the dimer is initially prepared in a coherent
superposition of vibronic exciton states provided that the chromophore-chromophore dipole interaction
is finite.

For vanishing chromophore-chromophore dipole interaction, the electronic 
and vibrational contributions to the vibronic Hamiltonian in Eq.~(\ref{equ:system_hamiltonian}) 
commute; thus indicating that a product state of electronic and vibrational single eigenstates
will also be an eigenstate --not a coherent superposition of eigenstates-- of the 
vibronic Hamiltonian. 
The fact that for this product state the Mandel parameter adopts only positive values led to 
conclude \cite{OReillyOlaya-Castro2014NC} that the transient formation of vibronic
exciton states 
establishes non-classical correlations in the vibrational modes.
However, the chromophore-chromophore dipole interaction is finite under sunlight illumination 
but not non-classical correlations are established provided the lack quantum correlations
in the initial state.
Therefore, non-classical correlations does not emerge due to the transient formation of 
vibronic exciton states, but as a consequence of the initial quantum correlations
established in the
light-harvesting system by the pulsed-laser-preparation of the initial state.

\section{Summary}
The role of intramolecular vibrations resonant with excitonic transitions in light-haversting systems
was analyzed under natural sunlight illumination.
The comprehensive analysis shows that the populations of single exciton and site states 
of vibronic dimers are not significantly affected as compared to their corresponding electronic 
dimers. 
Therefore, there is no direct evidence of an enhancement in the energy transport mediated by 
the inclusion of resonant intramolecular vibrational degrees of freedom under natural conditions.
Recently, similar conclusions were elucidated about the impact of the vibronic coupling 
in the electronic and vibrational coherences observed in two-dimensional-electronic-spectroscopies 
\cite{DuanThorwartMiller2019arXiv}.
However, the inclusion of the intramolecular vibrational modes reinforces the exciton 
coherence by up to one order of magnitude, as was shown for the DBV dimer.

Under incoherent light excitation conditions, it was further shown that intramolecular vibrational 
modes do not develop non-classical correlations provided that the initial state is devoid of 
quantum coherence. 
Therefore, the generation of non-classical correlations via the transient formation of
vibronic exciton states lacks of theoretical support and should be replaced in favor of
the natural dynamics of
the initial quantum correlations established in the light-harvesting system by the 
pulsed-laser-preparation of the initial state.

\begin{acknowledgments}
Broadband and chaotically enlighting discussions with Prof. Paul Brumer on excitation with 
incoherent light  are acknowledged with pleasure.
L.F.C. acknowledges financial support from the \emph{Departamento Administrativo de
Ciencia, Tecnolog\'ia e Innovaci\'on} --COLCIENCIAS--. 
This work was supported by the \emph{Comit\'e para el Desarrollo de la Investigaci\'on}
--CODI-- of Universidad de Antioquia, Colombia under the grant number 2015-7631 and 
by the \emph{Departamento Administrativo de Ciencia, Tecnolog\'ia e Innovaci\'on} 
--COLCIENCIAS-- of Colombia under the grant number 111556934912.
\end{acknowledgments}

\bibliography{IENAD}

\begin{thebibliography}{47}%
\makeatletter
\providecommand \@ifxundefined [1]{%
 \@ifx{#1\undefined}
}%
\providecommand \@ifnum [1]{%
 \ifnum #1\expandafter \@firstoftwo
 \else \expandafter \@secondoftwo
 \fi
}%
\providecommand \@ifx [1]{%
 \ifx #1\expandafter \@firstoftwo
 \else \expandafter \@secondoftwo
 \fi
}%
\providecommand \natexlab [1]{#1}%
\providecommand \enquote  [1]{``#1''}%
\providecommand \bibnamefont  [1]{#1}%
\providecommand \bibfnamefont [1]{#1}%
\providecommand \citenamefont [1]{#1}%
\providecommand \href@noop [0]{\@secondoftwo}%
\providecommand \href [0]{\begingroup \@sanitize@url \@href}%
\providecommand \@href[1]{\@@startlink{#1}\@@href}%
\providecommand \@@href[1]{\endgroup#1\@@endlink}%
\providecommand \@sanitize@url [0]{\catcode `\\12\catcode `\$12\catcode
  `\&12\catcode `\#12\catcode `\^12\catcode `\_12\catcode `\%12\relax}%
\providecommand \@@startlink[1]{}%
\providecommand \@@endlink[0]{}%
\providecommand \url  [0]{\begingroup\@sanitize@url \@url }%
\providecommand \@url [1]{\endgroup\@href {#1}{\urlprefix }}%
\providecommand \urlprefix  [0]{URL }%
\providecommand \Eprint [0]{\href }%
\providecommand \doibase [0]{http://dx.doi.org/}%
\providecommand \selectlanguage [0]{\@gobble}%
\providecommand \bibinfo  [0]{\@secondoftwo}%
\providecommand \bibfield  [0]{\@secondoftwo}%
\providecommand \translation [1]{[#1]}%
\providecommand \BibitemOpen [0]{}%
\providecommand \bibitemStop [0]{}%
\providecommand \bibitemNoStop [0]{.\EOS\space}%
\providecommand \EOS [0]{\spacefactor3000\relax}%
\providecommand \BibitemShut  [1]{\csname bibitem#1\endcsname}%
\let\auto@bib@innerbib\@empty
\bibitem [{\citenamefont {Pach\'{o}n}\ and\ \citenamefont
  {Brumer}(2012)}]{PachonBrumer2012PCCP}%
  \BibitemOpen
  \bibfield  {author} {\bibinfo {author} {\bibfnamefont {L.~A.}\ \bibnamefont
  {Pach\'{o}n}}\ and\ \bibinfo {author} {\bibfnamefont {P.}~\bibnamefont
  {Brumer}},\ }\href@noop {} {\bibfield  {journal} {\bibinfo  {journal} {Phys.
  Chem. Chem. Phys.}\ }\textbf {\bibinfo {volume} {14}},\ \bibinfo {pages}
  {10094} (\bibinfo {year} {2012})}\BibitemShut {NoStop}%
\bibitem [{\citenamefont {Chenu}\ and\ \citenamefont
  {Scholes}(2015)}]{ChenuScholes2015ARPC}%
  \BibitemOpen
  \bibfield  {author} {\bibinfo {author} {\bibfnamefont {A.}~\bibnamefont
  {Chenu}}\ and\ \bibinfo {author} {\bibfnamefont {G.~D.}\ \bibnamefont
  {Scholes}},\ }\href@noop {} {\bibfield  {journal} {\bibinfo  {journal} {Annu.
  Rev. Phys. Chem.}\ }\textbf {\bibinfo {volume} {66}},\ \bibinfo {pages} {69}
  (\bibinfo {year} {2015})}\BibitemShut {NoStop}%
\bibitem [{\citenamefont {Brunk}\ and\ \citenamefont
  {Rothlisberger}(2015)}]{BrunkRothlisberger2015CR}%
  \BibitemOpen
  \bibfield  {author} {\bibinfo {author} {\bibfnamefont {E.}~\bibnamefont
  {Brunk}}\ and\ \bibinfo {author} {\bibfnamefont {U.}~\bibnamefont
  {Rothlisberger}},\ }\href@noop {} {\bibfield  {journal} {\bibinfo  {journal}
  {Chem. Rev.}\ }\textbf {\bibinfo {volume} {115}},\ \bibinfo {pages} {6217}
  (\bibinfo {year} {2015})}\BibitemShut {NoStop}%
\bibitem [{\citenamefont {Curutchet}\ and\ \citenamefont
  {Mennucci}(2016)}]{CurutchetMennucci2016CR}%
  \BibitemOpen
  \bibfield  {author} {\bibinfo {author} {\bibfnamefont {C.}~\bibnamefont
  {Curutchet}}\ and\ \bibinfo {author} {\bibfnamefont {B.}~\bibnamefont
  {Mennucci}},\ }\href@noop {} {\bibfield  {journal} {\bibinfo  {journal}
  {Chem. Rev.}\ }\textbf {\bibinfo {volume} {117}},\ \bibinfo {pages} {294}
  (\bibinfo {year} {2016})}\BibitemShut {NoStop}%
\bibitem [{\citenamefont {de~Vega}\ and\ \citenamefont
  {Alonso}(2017)}]{deVegaAlonso2017RMP}%
  \BibitemOpen
  \bibfield  {author} {\bibinfo {author} {\bibfnamefont {I.}~\bibnamefont
  {de~Vega}}\ and\ \bibinfo {author} {\bibfnamefont {D.}~\bibnamefont
  {Alonso}},\ }\href@noop {} {\bibfield  {journal} {\bibinfo  {journal} {Rev.
  Mod. Phys.}\ }\textbf {\bibinfo {volume} {89}},\ \bibinfo {pages} {015001}
  (\bibinfo {year} {2017})}\BibitemShut {NoStop}%
\bibitem [{\citenamefont {Jang}\ and\ \citenamefont
  {Mennucci}(2018)}]{JangMenucci2018RMP}%
  \BibitemOpen
  \bibfield  {author} {\bibinfo {author} {\bibfnamefont {S.~J.}\ \bibnamefont
  {Jang}}\ and\ \bibinfo {author} {\bibfnamefont {B.}~\bibnamefont
  {Mennucci}},\ }\href@noop {} {\bibfield  {journal} {\bibinfo  {journal} {Rev.
  Mod. Phys.}\ }\textbf {\bibinfo {volume} {90}},\ \bibinfo {pages} {035003}
  (\bibinfo {year} {2018})}\BibitemShut {NoStop}%
\bibitem [{\citenamefont {Brumer}(2018)}]{Brumer2018JPCL}%
  \BibitemOpen
  \bibfield  {author} {\bibinfo {author} {\bibfnamefont {P.}~\bibnamefont
  {Brumer}},\ }\href@noop {} {\bibfield  {journal} {\bibinfo  {journal} {J.
  Phys. Chem. Lett.}\ }\textbf {\bibinfo {volume} {9}},\ \bibinfo {pages}
  {2946} (\bibinfo {year} {2018})}\BibitemShut {NoStop}%
\bibitem [{\citenamefont {Engel}\ \emph {et~al.}(2007)\citenamefont {Engel},
  \citenamefont {Calhoun}, \citenamefont {Read}, \citenamefont {Ahn},
  \citenamefont {Man{\v{c}}al}, \citenamefont {Cheng}, \citenamefont
  {Blankenship},\ and\ \citenamefont {Fleming}}]{EngelCalhounReadEtAl2007N}%
  \BibitemOpen
  \bibfield  {author} {\bibinfo {author} {\bibfnamefont {G.~S.}\ \bibnamefont
  {Engel}}, \bibinfo {author} {\bibfnamefont {T.~R.}\ \bibnamefont {Calhoun}},
  \bibinfo {author} {\bibfnamefont {E.~L.}\ \bibnamefont {Read}}, \bibinfo
  {author} {\bibfnamefont {T.-K.}\ \bibnamefont {Ahn}}, \bibinfo {author}
  {\bibfnamefont {T.}~\bibnamefont {Man{\v{c}}al}}, \bibinfo {author}
  {\bibfnamefont {Y.-C.}\ \bibnamefont {Cheng}}, \bibinfo {author}
  {\bibfnamefont {R.~E.}\ \bibnamefont {Blankenship}}, \ and\ \bibinfo {author}
  {\bibfnamefont {G.~R.}\ \bibnamefont {Fleming}},\ }\href@noop {} {\bibfield
  {journal} {\bibinfo  {journal} {Nature}\ }\textbf {\bibinfo {volume} {446}},\
  \bibinfo {pages} {782} (\bibinfo {year} {2007})}\BibitemShut {NoStop}%
\bibitem [{\citenamefont {Collini}\ \emph {et~al.}(2010)\citenamefont
  {Collini}, \citenamefont {Wong}, \citenamefont {Wilk}, \citenamefont {Curmi},
  \citenamefont {Brumer},\ and\ \citenamefont
  {Scholes}}]{ColliniWongWilkEtAl2010N}%
  \BibitemOpen
  \bibfield  {author} {\bibinfo {author} {\bibfnamefont {E.}~\bibnamefont
  {Collini}}, \bibinfo {author} {\bibfnamefont {C.~Y.}\ \bibnamefont {Wong}},
  \bibinfo {author} {\bibfnamefont {K.~E.}\ \bibnamefont {Wilk}}, \bibinfo
  {author} {\bibfnamefont {P.~M.~G.}\ \bibnamefont {Curmi}}, \bibinfo {author}
  {\bibfnamefont {P.}~\bibnamefont {Brumer}}, \ and\ \bibinfo {author}
  {\bibfnamefont {G.~D.}\ \bibnamefont {Scholes}},\ }\href@noop {} {\bibfield
  {journal} {\bibinfo  {journal} {Nature}\ }\textbf {\bibinfo {volume} {463}},\
  \bibinfo {pages} {644} (\bibinfo {year} {2010})}\BibitemShut {NoStop}%
\bibitem [{\citenamefont {Panitchayangkoon}\ \emph {et~al.}(2010)\citenamefont
  {Panitchayangkoon}, \citenamefont {Hayes}, \citenamefont {Fransted},
  \citenamefont {Caram}, \citenamefont {Harel}, \citenamefont {Wen},
  \citenamefont {Blankenship},\ and\ \citenamefont
  {Engel}}]{PanitchayangkoonHayesFranstedEtAl2010PNAS}%
  \BibitemOpen
  \bibfield  {author} {\bibinfo {author} {\bibfnamefont {G.}~\bibnamefont
  {Panitchayangkoon}}, \bibinfo {author} {\bibfnamefont {D.}~\bibnamefont
  {Hayes}}, \bibinfo {author} {\bibfnamefont {K.~A.}\ \bibnamefont {Fransted}},
  \bibinfo {author} {\bibfnamefont {J.~R.}\ \bibnamefont {Caram}}, \bibinfo
  {author} {\bibfnamefont {E.}~\bibnamefont {Harel}}, \bibinfo {author}
  {\bibfnamefont {J.}~\bibnamefont {Wen}}, \bibinfo {author} {\bibfnamefont
  {R.~E.}\ \bibnamefont {Blankenship}}, \ and\ \bibinfo {author} {\bibfnamefont
  {G.~S.}\ \bibnamefont {Engel}},\ }\href@noop {} {\bibfield  {journal}
  {\bibinfo  {journal} {Proc. Natl. Acad. Sci. U.S.A.}\ }\textbf {\bibinfo
  {volume} {107}},\ \bibinfo {pages} {12766} (\bibinfo {year}
  {2010})}\BibitemShut {NoStop}%
\bibitem [{\citenamefont {Ishizaki}\ \emph {et~al.}(2010)\citenamefont
  {Ishizaki}, \citenamefont {Calhoun}, \citenamefont {Schlau-Cohen},\ and\
  \citenamefont {Fleming}}]{IshizakiCalhounSchlau-CohenEtAl2010PCCP}%
  \BibitemOpen
  \bibfield  {author} {\bibinfo {author} {\bibfnamefont {A.}~\bibnamefont
  {Ishizaki}}, \bibinfo {author} {\bibfnamefont {T.~R.}\ \bibnamefont
  {Calhoun}}, \bibinfo {author} {\bibfnamefont {G.~S.}\ \bibnamefont
  {Schlau-Cohen}}, \ and\ \bibinfo {author} {\bibfnamefont {G.~R.}\
  \bibnamefont {Fleming}},\ }\href@noop {} {\bibfield  {journal} {\bibinfo
  {journal} {Phys. Chem. Chem. Phys.}\ }\textbf {\bibinfo {volume} {12}},\
  \bibinfo {pages} {7319} (\bibinfo {year} {2010})}\BibitemShut {NoStop}%
\bibitem [{\citenamefont {Pach\'{o}n}\ and\ \citenamefont
  {Brumer}(2011)}]{PachonBrumer2011JPCL}%
  \BibitemOpen
  \bibfield  {author} {\bibinfo {author} {\bibfnamefont {L.~A.}\ \bibnamefont
  {Pach\'{o}n}}\ and\ \bibinfo {author} {\bibfnamefont {P.}~\bibnamefont
  {Brumer}},\ }\href@noop {} {\bibfield  {journal} {\bibinfo  {journal} {J.
  Phys. Chem. Lett.}\ }\textbf {\bibinfo {volume} {2}},\ \bibinfo {pages}
  {2728} (\bibinfo {year} {2011})}\BibitemShut {NoStop}%
\bibitem [{\citenamefont {Huelga}\ and\ \citenamefont
  {Plenio}(2013)}]{HuelgaPlenio2013CP}%
  \BibitemOpen
  \bibfield  {author} {\bibinfo {author} {\bibfnamefont {S.~F.}\ \bibnamefont
  {Huelga}}\ and\ \bibinfo {author} {\bibfnamefont {M.~B.}\ \bibnamefont
  {Plenio}},\ }\href@noop {} {\bibfield  {journal} {\bibinfo  {journal}
  {Contemp. Phys.}\ }\textbf {\bibinfo {volume} {54}},\ \bibinfo {pages} {181}
  (\bibinfo {year} {2013})}\BibitemShut {NoStop}%
\bibitem [{\citenamefont {Christensson}\ \emph {et~al.}(2012)\citenamefont
  {Christensson}, \citenamefont {Kauffmann}, \citenamefont {Pullerits},\ and\
  \citenamefont {Man{\v c}al}}]{ChristenssonKauffmannPulleritsEtAl2012JPCB}%
  \BibitemOpen
  \bibfield  {author} {\bibinfo {author} {\bibfnamefont {N.}~\bibnamefont
  {Christensson}}, \bibinfo {author} {\bibfnamefont {H.~F.}\ \bibnamefont
  {Kauffmann}}, \bibinfo {author} {\bibfnamefont {T.}~\bibnamefont
  {Pullerits}}, \ and\ \bibinfo {author} {\bibfnamefont {T.}~\bibnamefont
  {Man{\v c}al}},\ }\href@noop {} {\bibfield  {journal} {\bibinfo  {journal}
  {J. Phys. Chem. B}\ }\textbf {\bibinfo {volume} {116}},\ \bibinfo {pages}
  {7449} (\bibinfo {year} {2012})}\BibitemShut {NoStop}%
\bibitem [{\citenamefont {Kolli}\ \emph {et~al.}(2012)\citenamefont {Kolli},
  \citenamefont {O'Reilly}, \citenamefont {Scholes},\ and\ \citenamefont
  {Olaya-Castro}}]{KolliOReillyScholesEtAl2012JCP}%
  \BibitemOpen
  \bibfield  {author} {\bibinfo {author} {\bibfnamefont {A.}~\bibnamefont
  {Kolli}}, \bibinfo {author} {\bibfnamefont {E.~J.}\ \bibnamefont {O'Reilly}},
  \bibinfo {author} {\bibfnamefont {G.~D.}\ \bibnamefont {Scholes}}, \ and\
  \bibinfo {author} {\bibfnamefont {A.}~\bibnamefont {Olaya-Castro}},\
  }\href@noop {} {\bibfield  {journal} {\bibinfo  {journal} {J. Chem. Phys.}\
  }\textbf {\bibinfo {volume} {137}},\ \bibinfo {pages} {174109} (\bibinfo
  {year} {2012})}\BibitemShut {NoStop}%
\bibitem [{\citenamefont {Tiwari}\ \emph {et~al.}(2013)\citenamefont {Tiwari},
  \citenamefont {Peters},\ and\ \citenamefont
  {Jonas}}]{TiwariPetersJonas2013PNASU}%
  \BibitemOpen
  \bibfield  {author} {\bibinfo {author} {\bibfnamefont {V.}~\bibnamefont
  {Tiwari}}, \bibinfo {author} {\bibfnamefont {W.~K.}\ \bibnamefont {Peters}},
  \ and\ \bibinfo {author} {\bibfnamefont {D.~M.}\ \bibnamefont {Jonas}},\
  }\href@noop {} {\bibfield  {journal} {\bibinfo  {journal} {Proc. Natl. Acad.
  Sci. USA}\ }\textbf {\bibinfo {volume} {110}},\ \bibinfo {pages} {1203}
  (\bibinfo {year} {2013})}\BibitemShut {NoStop}%
\bibitem [{\citenamefont {Chin}\ \emph {et~al.}(2013)\citenamefont {Chin},
  \citenamefont {Prior}, \citenamefont {Rosenbach}, \citenamefont
  {Caycedo-Soler}, \citenamefont {Huelga},\ and\ \citenamefont
  {Plenio}}]{ChinPriorRosenbachEtAl2013NP}%
  \BibitemOpen
  \bibfield  {author} {\bibinfo {author} {\bibfnamefont {A.~W.}\ \bibnamefont
  {Chin}}, \bibinfo {author} {\bibfnamefont {J.}~\bibnamefont {Prior}},
  \bibinfo {author} {\bibfnamefont {R.}~\bibnamefont {Rosenbach}}, \bibinfo
  {author} {\bibfnamefont {F.}~\bibnamefont {Caycedo-Soler}}, \bibinfo {author}
  {\bibfnamefont {S.~F.}\ \bibnamefont {Huelga}}, \ and\ \bibinfo {author}
  {\bibfnamefont {M.~B.}\ \bibnamefont {Plenio}},\ }\href@noop {} {\bibfield
  {journal} {\bibinfo  {journal} {Nat. Phys.}\ }\textbf {\bibinfo {volume}
  {9}},\ \bibinfo {pages} {113} (\bibinfo {year} {2013})}\BibitemShut {NoStop}%
\bibitem [{\citenamefont {Chenu}\ \emph {et~al.}(2013)\citenamefont {Chenu},
  \citenamefont {Christensson}, \citenamefont {Kauffmann},\ and\ \citenamefont
  {Man{\v{c}}al}}]{ChenuChristenssonKauffmannEtAl2013SR}%
  \BibitemOpen
  \bibfield  {author} {\bibinfo {author} {\bibfnamefont {A.}~\bibnamefont
  {Chenu}}, \bibinfo {author} {\bibfnamefont {N.}~\bibnamefont {Christensson}},
  \bibinfo {author} {\bibfnamefont {H.~F.}\ \bibnamefont {Kauffmann}}, \ and\
  \bibinfo {author} {\bibfnamefont {T.}~\bibnamefont {Man{\v{c}}al}},\
  }\href@noop {} {\bibfield  {journal} {\bibinfo  {journal} {Sci. Rep.}\
  }\textbf {\bibinfo {volume} {3}},\ \bibinfo {pages} {2029} (\bibinfo {year}
  {2013})}\BibitemShut {NoStop}%
\bibitem [{\citenamefont {Novelli}\ \emph {et~al.}(2015)\citenamefont
  {Novelli}, \citenamefont {Nazir}, \citenamefont {Richards}, \citenamefont
  {Roozbeh}, \citenamefont {Wilk}, \citenamefont {Curmi},\ and\ \citenamefont
  {Davis}}]{NovelliNazirRichardsEtAl2015JPCL}%
  \BibitemOpen
  \bibfield  {author} {\bibinfo {author} {\bibfnamefont {F.}~\bibnamefont
  {Novelli}}, \bibinfo {author} {\bibfnamefont {A.}~\bibnamefont {Nazir}},
  \bibinfo {author} {\bibfnamefont {G.~H.}\ \bibnamefont {Richards}}, \bibinfo
  {author} {\bibfnamefont {A.}~\bibnamefont {Roozbeh}}, \bibinfo {author}
  {\bibfnamefont {K.~E.}\ \bibnamefont {Wilk}}, \bibinfo {author}
  {\bibfnamefont {P.~M.}\ \bibnamefont {Curmi}}, \ and\ \bibinfo {author}
  {\bibfnamefont {J.~A.}\ \bibnamefont {Davis}},\ }\href@noop {} {\bibfield
  {journal} {\bibinfo  {journal} {J. Phys. Chem. Lett.}\ }\textbf {\bibinfo
  {volume} {6}},\ \bibinfo {pages} {4573} (\bibinfo {year} {2015})}\BibitemShut
  {NoStop}%
\bibitem [{\citenamefont {Mal\'{y}}\ \emph {et~al.}(2016)\citenamefont
  {Mal\'{y}}, \citenamefont {Somsen}, \citenamefont {Novoderezhkin},
  \citenamefont {Man\v{c}al},\ and\ \citenamefont
  {Van~Grondelle}}]{MalySomsenNovoderezhkinEtAl2016CPC}%
  \BibitemOpen
  \bibfield  {author} {\bibinfo {author} {\bibfnamefont {P.}~\bibnamefont
  {Mal\'{y}}}, \bibinfo {author} {\bibfnamefont {O.~J.}\ \bibnamefont
  {Somsen}}, \bibinfo {author} {\bibfnamefont {V.~I.}\ \bibnamefont
  {Novoderezhkin}}, \bibinfo {author} {\bibfnamefont {T.}~\bibnamefont
  {Man\v{c}al}}, \ and\ \bibinfo {author} {\bibfnamefont {R.}~\bibnamefont
  {Van~Grondelle}},\ }\href@noop {} {\bibfield  {journal} {\bibinfo  {journal}
  {ChemPhysChem}\ }\textbf {\bibinfo {volume} {17}},\ \bibinfo {pages} {1356}
  (\bibinfo {year} {2016})}\BibitemShut {NoStop}%
\bibitem [{\citenamefont {Dean}\ \emph {et~al.}(2016)\citenamefont {Dean},
  \citenamefont {Mirkovic}, \citenamefont {Toa}, \citenamefont {Oblinsky},\
  and\ \citenamefont {Scholes}}]{DeanMirkovicToaEtAl2016C}%
  \BibitemOpen
  \bibfield  {author} {\bibinfo {author} {\bibfnamefont {J.~C.}\ \bibnamefont
  {Dean}}, \bibinfo {author} {\bibfnamefont {T.}~\bibnamefont {Mirkovic}},
  \bibinfo {author} {\bibfnamefont {Z.~S.}\ \bibnamefont {Toa}}, \bibinfo
  {author} {\bibfnamefont {D.~G.}\ \bibnamefont {Oblinsky}}, \ and\ \bibinfo
  {author} {\bibfnamefont {G.~D.}\ \bibnamefont {Scholes}},\ }\href@noop {}
  {\bibfield  {journal} {\bibinfo  {journal} {Chem}\ }\textbf {\bibinfo
  {volume} {1}},\ \bibinfo {pages} {858} (\bibinfo {year} {2016})}\BibitemShut
  {NoStop}%
\bibitem [{\citenamefont {Yeh}\ \emph {et~al.}(2018)\citenamefont {Yeh},
  \citenamefont {Hoehn}, \citenamefont {Allodi}, \citenamefont {Engel},\ and\
  \citenamefont {Kais}}]{YehHoehnAllodiEtAl2018PNAS}%
  \BibitemOpen
  \bibfield  {author} {\bibinfo {author} {\bibfnamefont {S.-H.}\ \bibnamefont
  {Yeh}}, \bibinfo {author} {\bibfnamefont {R.~D.}\ \bibnamefont {Hoehn}},
  \bibinfo {author} {\bibfnamefont {M.~A.}\ \bibnamefont {Allodi}}, \bibinfo
  {author} {\bibfnamefont {G.~S.}\ \bibnamefont {Engel}}, \ and\ \bibinfo
  {author} {\bibfnamefont {S.}~\bibnamefont {Kais}},\ }\href@noop {} {\bibfield
   {journal} {\bibinfo  {journal} {Proc. Natl. Acad. Sci. U.S.A.}\ ,\ \bibinfo
  {pages} {201701390}} (\bibinfo {year} {2018})}\BibitemShut {NoStop}%
\bibitem [{\citenamefont {O'Reilly}\ and\ \citenamefont
  {Olaya-Castro}(2014)}]{OReillyOlaya-Castro2014NC}%
  \BibitemOpen
  \bibfield  {author} {\bibinfo {author} {\bibfnamefont {E.~J.}\ \bibnamefont
  {O'Reilly}}\ and\ \bibinfo {author} {\bibfnamefont {A.}~\bibnamefont
  {Olaya-Castro}},\ }\href@noop {} {\bibfield  {journal} {\bibinfo  {journal}
  {Nat. Commun.}\ }\textbf {\bibinfo {volume} {5}} (\bibinfo {year}
  {2014})}\BibitemShut {NoStop}%
\bibitem [{\citenamefont {Mukamel}(1995)}]{Mukamel:1995}%
  \BibitemOpen
  \bibfield  {author} {\bibinfo {author} {\bibfnamefont {S.}~\bibnamefont
  {Mukamel}},\ }\href@noop {} {\emph {\bibinfo {title} {Principles of Nonlinear
  Spectroscopy}}}\ (\bibinfo  {publisher} {Oxford University Press},\ \bibinfo
  {year} {1995})\BibitemShut {NoStop}%
\bibitem [{\citenamefont {Man{\v{c}}al}\ and\ \citenamefont
  {Valkunas}(2010)}]{MancalValkunas2010NJP}%
  \BibitemOpen
  \bibfield  {author} {\bibinfo {author} {\bibfnamefont {T.}~\bibnamefont
  {Man{\v{c}}al}}\ and\ \bibinfo {author} {\bibfnamefont {L.}~\bibnamefont
  {Valkunas}},\ }\href@noop {} {\bibfield  {journal} {\bibinfo  {journal} {New
  J. Phys.}\ }\textbf {\bibinfo {volume} {12}},\ \bibinfo {pages} {065044}
  (\bibinfo {year} {2010})}\BibitemShut {NoStop}%
\bibitem [{\citenamefont {Brumer}\ and\ \citenamefont
  {Shapiro}(2012)}]{BrumerShapiro2012PNASU}%
  \BibitemOpen
  \bibfield  {author} {\bibinfo {author} {\bibfnamefont {P.}~\bibnamefont
  {Brumer}}\ and\ \bibinfo {author} {\bibfnamefont {M.}~\bibnamefont
  {Shapiro}},\ }\href@noop {} {\bibfield  {journal} {\bibinfo  {journal} {Proc.
  Natl. Acad. Sci. U.S.A.}\ }\textbf {\bibinfo {volume} {109}},\ \bibinfo
  {pages} {19575} (\bibinfo {year} {2012})}\BibitemShut {NoStop}%
\bibitem [{\citenamefont {Tscherbul}\ and\ \citenamefont
  {Brumer}(2014)}]{TscherbulBrumer2014PRL}%
  \BibitemOpen
  \bibfield  {author} {\bibinfo {author} {\bibfnamefont {T.~V.}\ \bibnamefont
  {Tscherbul}}\ and\ \bibinfo {author} {\bibfnamefont {P.}~\bibnamefont
  {Brumer}},\ }\href@noop {} {\bibfield  {journal} {\bibinfo  {journal} {Phys.
  Rev. Lett.}\ }\textbf {\bibinfo {volume} {113}},\ \bibinfo {pages} {113601}
  (\bibinfo {year} {2014})}\BibitemShut {NoStop}%
\bibitem [{\citenamefont {Sadeq}\ and\ \citenamefont
  {Brumer}(2014)}]{SadeqBrumer2014JCP}%
  \BibitemOpen
  \bibfield  {author} {\bibinfo {author} {\bibfnamefont {Z.~S.}\ \bibnamefont
  {Sadeq}}\ and\ \bibinfo {author} {\bibfnamefont {P.}~\bibnamefont {Brumer}},\
  }\href@noop {} {\bibfield  {journal} {\bibinfo  {journal} {J. Chem. Phys.}\
  }\textbf {\bibinfo {volume} {140}},\ \bibinfo {pages} {074104} (\bibinfo
  {year} {2014})}\BibitemShut {NoStop}%
\bibitem [{\citenamefont {Grinev}\ and\ \citenamefont
  {Brumer}(2015)}]{GrinevBrumer2015JCP}%
  \BibitemOpen
  \bibfield  {author} {\bibinfo {author} {\bibfnamefont {T.}~\bibnamefont
  {Grinev}}\ and\ \bibinfo {author} {\bibfnamefont {P.}~\bibnamefont
  {Brumer}},\ }\href@noop {} {\bibfield  {journal} {\bibinfo  {journal} {J.
  Chem. Phys.}\ }\textbf {\bibinfo {volume} {143}},\ \bibinfo {pages} {244313}
  (\bibinfo {year} {2015})}\BibitemShut {NoStop}%
\bibitem [{\citenamefont {Dodin}\ \emph {et~al.}(2016)\citenamefont {Dodin},
  \citenamefont {Tscherbul},\ and\ \citenamefont
  {Brumer}}]{DodinTscherbulBrumer2016JCP}%
  \BibitemOpen
  \bibfield  {author} {\bibinfo {author} {\bibfnamefont {A.}~\bibnamefont
  {Dodin}}, \bibinfo {author} {\bibfnamefont {T.~V.}\ \bibnamefont
  {Tscherbul}}, \ and\ \bibinfo {author} {\bibfnamefont {P.}~\bibnamefont
  {Brumer}},\ }\href@noop {} {\bibfield  {journal} {\bibinfo  {journal} {J.
  Chem. Phys.}\ }\textbf {\bibinfo {volume} {144}},\ \bibinfo {pages} {244108}
  (\bibinfo {year} {2016})}\BibitemShut {NoStop}%
\bibitem [{\citenamefont {Pach{\'o}n}\ \emph {et~al.}(2017)\citenamefont
  {Pach{\'o}n}, \citenamefont {Botero},\ and\ \citenamefont
  {Brumer}}]{PachonBoteroBrumer2017JPBAMOP}%
  \BibitemOpen
  \bibfield  {author} {\bibinfo {author} {\bibfnamefont {L.~A.}\ \bibnamefont
  {Pach{\'o}n}}, \bibinfo {author} {\bibfnamefont {J.~D.}\ \bibnamefont
  {Botero}}, \ and\ \bibinfo {author} {\bibfnamefont {P.~W.}\ \bibnamefont
  {Brumer}},\ }\href@noop {} {\bibfield  {journal} {\bibinfo  {journal} {J.
  Phys. B: At. Mol. Opt. Phys.}\ }\textbf {\bibinfo {volume} {50}},\ \bibinfo
  {pages} {184003} (\bibinfo {year} {2017})}\BibitemShut {NoStop}%
\bibitem [{\citenamefont {Chenu}\ \emph {et~al.}(2014)\citenamefont {Chenu},
  \citenamefont {Mal{\`y}},\ and\ \citenamefont
  {Man{\v{c}}al}}]{ChenuMalyMancal2014CP}%
  \BibitemOpen
  \bibfield  {author} {\bibinfo {author} {\bibfnamefont {A.}~\bibnamefont
  {Chenu}}, \bibinfo {author} {\bibfnamefont {P.}~\bibnamefont {Mal{\`y}}}, \
  and\ \bibinfo {author} {\bibfnamefont {T.}~\bibnamefont {Man{\v{c}}al}},\
  }\href@noop {} {\bibfield  {journal} {\bibinfo  {journal} {Chem. Phys.}\
  }\textbf {\bibinfo {volume} {439}},\ \bibinfo {pages} {100} (\bibinfo {year}
  {2014})}\BibitemShut {NoStop}%
\bibitem [{\citenamefont {Chenu}\ \emph {et~al.}(2015)\citenamefont {Chenu},
  \citenamefont {Bra\'{n}czyk}, \citenamefont {Scholes},\ and\ \citenamefont
  {Sipe}}]{ChenuBranczykScholesEtAl2015PRL}%
  \BibitemOpen
  \bibfield  {author} {\bibinfo {author} {\bibfnamefont {A.}~\bibnamefont
  {Chenu}}, \bibinfo {author} {\bibfnamefont {A.~M.}\ \bibnamefont
  {Bra\'{n}czyk}}, \bibinfo {author} {\bibfnamefont {G.~D.}\ \bibnamefont
  {Scholes}}, \ and\ \bibinfo {author} {\bibfnamefont {J.~E.}\ \bibnamefont
  {Sipe}},\ }\href@noop {} {\bibfield  {journal} {\bibinfo  {journal} {Phys.
  Rev. Lett.}\ }\textbf {\bibinfo {volume} {114}},\ \bibinfo {pages} {213601}
  (\bibinfo {year} {2015})}\BibitemShut {NoStop}%
\bibitem [{\citenamefont {Chenu}\ and\ \citenamefont
  {Brumer}(2016)}]{ChenuBrumer2016JCP}%
  \BibitemOpen
  \bibfield  {author} {\bibinfo {author} {\bibfnamefont {A.}~\bibnamefont
  {Chenu}}\ and\ \bibinfo {author} {\bibfnamefont {P.}~\bibnamefont {Brumer}},\
  }\href@noop {} {\bibfield  {journal} {\bibinfo  {journal} {J. Chem. Phys.}\
  }\textbf {\bibinfo {volume} {144}},\ \bibinfo {pages} {044103} (\bibinfo
  {year} {2016})}\BibitemShut {NoStop}%
\bibitem [{\citenamefont {Pach\'{o}n}\ and\ \citenamefont
  {Brumer}(2013)}]{PachonBrumer2013PRA}%
  \BibitemOpen
  \bibfield  {author} {\bibinfo {author} {\bibfnamefont {L.~A.}\ \bibnamefont
  {Pach\'{o}n}}\ and\ \bibinfo {author} {\bibfnamefont {P.}~\bibnamefont
  {Brumer}},\ }\href@noop {} {\bibfield  {journal} {\bibinfo  {journal} {Phys.
  Rev. A}\ }\textbf {\bibinfo {volume} {87}},\ \bibinfo {pages} {022106}
  (\bibinfo {year} {2013})}\BibitemShut {NoStop}%
\bibitem [{\citenamefont {Scholes}\ \emph {et~al.}(2017)\citenamefont
  {Scholes}, \citenamefont {Fleming}, \citenamefont {Chen}, \citenamefont
  {Aspuru-Guzik}, \citenamefont {Buchleitner}, \citenamefont {Coker},
  \citenamefont {Engel}, \citenamefont {van Grondelle}, \citenamefont
  {Ishizaki}, \citenamefont {Jonas},\ and\ \citenamefont
  {et~al.}}]{ScholesFlemingChen2017N}%
  \BibitemOpen
  \bibfield  {author} {\bibinfo {author} {\bibfnamefont {G.~D.}\ \bibnamefont
  {Scholes}}, \bibinfo {author} {\bibfnamefont {G.~R.}\ \bibnamefont
  {Fleming}}, \bibinfo {author} {\bibfnamefont {L.~X.}\ \bibnamefont {Chen}},
  \bibinfo {author} {\bibfnamefont {A.}~\bibnamefont {Aspuru-Guzik}}, \bibinfo
  {author} {\bibfnamefont {A.}~\bibnamefont {Buchleitner}}, \bibinfo {author}
  {\bibfnamefont {D.~F.}\ \bibnamefont {Coker}}, \bibinfo {author}
  {\bibfnamefont {G.~S.}\ \bibnamefont {Engel}}, \bibinfo {author}
  {\bibfnamefont {R.}~\bibnamefont {van Grondelle}}, \bibinfo {author}
  {\bibfnamefont {A.}~\bibnamefont {Ishizaki}}, \bibinfo {author}
  {\bibfnamefont {D.~M.}\ \bibnamefont {Jonas}}, \ and\ \bibinfo {author}
  {\bibnamefont {et~al.}},\ }\href@noop {} {\bibfield  {journal} {\bibinfo
  {journal} {Nature}\ }\textbf {\bibinfo {volume} {543}},\ \bibinfo {pages}
  {647} (\bibinfo {year} {2017})}\BibitemShut {NoStop}%
\bibitem [{\citenamefont {Mandel}\ and\ \citenamefont
  {Wolf}(1995)}]{Mandel-Wolf:1995}%
  \BibitemOpen
  \bibfield  {author} {\bibinfo {author} {\bibfnamefont {L.}~\bibnamefont
  {Mandel}}\ and\ \bibinfo {author} {\bibfnamefont {E.}~\bibnamefont {Wolf}},\
  }\href@noop {} {\emph {\bibinfo {title} {Optical coherence and quantum
  optics}}}\ (\bibinfo  {publisher} {Cambridge University Press},\ \bibinfo
  {year} {1995})\BibitemShut {NoStop}%
\bibitem [{\citenamefont {Tanimura}\ and\ \citenamefont
  {Kubo}(1989)}]{TanimuraKubo1989JPSJ}%
  \BibitemOpen
  \bibfield  {author} {\bibinfo {author} {\bibfnamefont {Y.}~\bibnamefont
  {Tanimura}}\ and\ \bibinfo {author} {\bibfnamefont {R.}~\bibnamefont
  {Kubo}},\ }\href@noop {} {\bibfield  {journal} {\bibinfo  {journal} {J. Phys.
  Soc. Jpn.}\ }\textbf {\bibinfo {volume} {58}},\ \bibinfo {pages} {101}
  (\bibinfo {year} {1989})}\BibitemShut {NoStop}%
\bibitem [{\citenamefont {Ishizaki}\ and\ \citenamefont
  {Tanimura}(2005)}]{IshizakiTanimura2005JPSJ}%
  \BibitemOpen
  \bibfield  {author} {\bibinfo {author} {\bibfnamefont {A.}~\bibnamefont
  {Ishizaki}}\ and\ \bibinfo {author} {\bibfnamefont {Y.}~\bibnamefont
  {Tanimura}},\ }\href@noop {} {\bibfield  {journal} {\bibinfo  {journal} {J.
  Phys. Soc. Jpn.}\ }\textbf {\bibinfo {volume} {74}},\ \bibinfo {pages} {3131}
  (\bibinfo {year} {2005})}\BibitemShut {NoStop}%
\bibitem [{\citenamefont {Ishizaki}\ and\ \citenamefont
  {Fleming}(2009)}]{IshizakiFleming2009JCP}%
  \BibitemOpen
  \bibfield  {author} {\bibinfo {author} {\bibfnamefont {A.}~\bibnamefont
  {Ishizaki}}\ and\ \bibinfo {author} {\bibfnamefont {G.~R.}\ \bibnamefont
  {Fleming}},\ }\href@noop {} {\bibfield  {journal} {\bibinfo  {journal} {J.
  Chem. Phys.}\ }\textbf {\bibinfo {volume} {130}},\ \bibinfo {pages} {234111}
  (\bibinfo {year} {2009})}\BibitemShut {NoStop}%
\bibitem [{\citenamefont {Fassioli}\ \emph {et~al.}(2012)\citenamefont
  {Fassioli}, \citenamefont {Olaya-Castro},\ and\ \citenamefont
  {Scholes}}]{FassioliOlaya-CastroScholes2012JPCL}%
  \BibitemOpen
  \bibfield  {author} {\bibinfo {author} {\bibfnamefont {F.}~\bibnamefont
  {Fassioli}}, \bibinfo {author} {\bibfnamefont {A.}~\bibnamefont
  {Olaya-Castro}}, \ and\ \bibinfo {author} {\bibfnamefont {G.~D.}\
  \bibnamefont {Scholes}},\ }\href@noop {} {\bibfield  {journal} {\bibinfo
  {journal} {J. Phys. Chem. Lett.}\ }\textbf {\bibinfo {volume} {3}},\ \bibinfo
  {pages} {3136} (\bibinfo {year} {2012})}\BibitemShut {NoStop}%
\bibitem [{\citenamefont {Chan}\ \emph {et~al.}(2018)\citenamefont {Chan},
  \citenamefont {Gamel}, \citenamefont {Fleming},\ and\ \citenamefont
  {Whaley}}]{ChanGamelFlemingEtAl2018JPB}%
  \BibitemOpen
  \bibfield  {author} {\bibinfo {author} {\bibfnamefont {H.~C.}\ \bibnamefont
  {Chan}}, \bibinfo {author} {\bibfnamefont {O.~E.}\ \bibnamefont {Gamel}},
  \bibinfo {author} {\bibfnamefont {G.~R.}\ \bibnamefont {Fleming}}, \ and\
  \bibinfo {author} {\bibfnamefont {K.~B.}\ \bibnamefont {Whaley}},\
  }\href@noop {} {\bibfield  {journal} {\bibinfo  {journal} {J. Phys. B}\
  }\textbf {\bibinfo {volume} {51}},\ \bibinfo {pages} {054002} (\bibinfo
  {year} {2018})}\BibitemShut {NoStop}%
\bibitem [{\citenamefont {May}\ and\ \citenamefont
  {K\"{u}hn}(2011)}]{May-Kuhn:2011}%
  \BibitemOpen
  \bibfield  {author} {\bibinfo {author} {\bibfnamefont {V.}~\bibnamefont
  {May}}\ and\ \bibinfo {author} {\bibfnamefont {O.}~\bibnamefont {K\"{u}hn}},\
  }\href@noop {} {\emph {\bibinfo {title} {Charge and Energy Transfer Dynamics
  in Molecular Systems}}}\ (\bibinfo  {publisher} {WILEY-VCH},\ \bibinfo {year}
  {2011})\BibitemShut {NoStop}%
\bibitem [{\citenamefont {Romero}\ \emph {et~al.}(2014)\citenamefont {Romero},
  \citenamefont {Augulis}, \citenamefont {Novoderezhkin}, \citenamefont
  {Ferretti}, \citenamefont {Thieme}, \citenamefont {Zigmantas},\ and\
  \citenamefont {Van~Grondelle}}]{RomeroAugulisNovoderezhkinEtAl2014NP}%
  \BibitemOpen
  \bibfield  {author} {\bibinfo {author} {\bibfnamefont {E.}~\bibnamefont
  {Romero}}, \bibinfo {author} {\bibfnamefont {R.}~\bibnamefont {Augulis}},
  \bibinfo {author} {\bibfnamefont {V.~I.}\ \bibnamefont {Novoderezhkin}},
  \bibinfo {author} {\bibfnamefont {M.}~\bibnamefont {Ferretti}}, \bibinfo
  {author} {\bibfnamefont {J.}~\bibnamefont {Thieme}}, \bibinfo {author}
  {\bibfnamefont {D.}~\bibnamefont {Zigmantas}}, \ and\ \bibinfo {author}
  {\bibfnamefont {R.}~\bibnamefont {Van~Grondelle}},\ }\href@noop {} {\bibfield
   {journal} {\bibinfo  {journal} {Nat. Phys.}\ }\textbf {\bibinfo {volume}
  {10}},\ \bibinfo {pages} {676} (\bibinfo {year} {2014})}\BibitemShut
  {NoStop}%
\bibitem [{\citenamefont {Novoderezhkin}\ and\ \citenamefont {van
  Grondelle}(2017)}]{NovoderezhkinGrondelle2017JPB}%
  \BibitemOpen
  \bibfield  {author} {\bibinfo {author} {\bibfnamefont {V.~I.}\ \bibnamefont
  {Novoderezhkin}}\ and\ \bibinfo {author} {\bibfnamefont {R.}~\bibnamefont
  {van Grondelle}},\ }\href@noop {} {\bibfield  {journal} {\bibinfo  {journal}
  {J. Phys. B}\ }\textbf {\bibinfo {volume} {50}},\ \bibinfo {pages} {124003}
  (\bibinfo {year} {2017})}\BibitemShut {NoStop}%
\bibitem [{\citenamefont {Bennett}\ \emph {et~al.}(2018)\citenamefont
  {Bennett}, \citenamefont {Maly}, \citenamefont {Kreisbeck}, \citenamefont
  {van Grondelle},\ and\ \citenamefont
  {Aspuru-Guzik}}]{BennettMalyKreisbeckEtAl2018JPCL}%
  \BibitemOpen
  \bibfield  {author} {\bibinfo {author} {\bibfnamefont {D.~I.}\ \bibnamefont
  {Bennett}}, \bibinfo {author} {\bibfnamefont {P.}~\bibnamefont {Maly}},
  \bibinfo {author} {\bibfnamefont {C.}~\bibnamefont {Kreisbeck}}, \bibinfo
  {author} {\bibfnamefont {R.}~\bibnamefont {van Grondelle}}, \ and\ \bibinfo
  {author} {\bibfnamefont {A.}~\bibnamefont {Aspuru-Guzik}},\ }\href@noop {}
  {\bibfield  {journal} {\bibinfo  {journal} {J. Phys. Chem. Lett.}\ }\textbf
  {\bibinfo {volume} {9}},\ \bibinfo {pages} {2665} (\bibinfo {year}
  {2018})}\BibitemShut {NoStop}%
\bibitem [{\citenamefont {Duan}\ \emph {et~al.}(2019)\citenamefont {Duan},
  \citenamefont {Thorwart},\ and\ \citenamefont
  {Miller}}]{DuanThorwartMiller2019arXiv}%
  \BibitemOpen
  \bibfield  {author} {\bibinfo {author} {\bibfnamefont {H.-G.}\ \bibnamefont
  {Duan}}, \bibinfo {author} {\bibfnamefont {M.}~\bibnamefont {Thorwart}}, \
  and\ \bibinfo {author} {\bibfnamefont {R.}~\bibnamefont {Miller}},\
  }\href@noop {} {\bibfield  {journal} {\bibinfo  {journal} {arXiv:1904.04033}\
  } (\bibinfo {year} {2019})}\BibitemShut {NoStop}%
\end{thebibliography}%

\end{document}